\documentclass[printer]{raa}  
\usepackage{natbib}
\usepackage{graphicx,times}             
\graphicspath{{figs}}
\usepackage{amssymb,amsmath,mathrsfs}
\usepackage{multirow}
\usepackage{booktabs,threeparttable}
\bibpunct{(}{)}{;}{a}{}{,}
\usepackage[pagebackref=true]{hyperref}
\usepackage{rotating}
\usepackage{enumitem}

\newcommand{\bea}{\begin{eqnarray}}
\newcommand{\eea}{\end{eqnarray}}

\usepackage{ulem}

\begin{document}

\title{Pixel-level modelling of group-scale strong lens CASSOWARY 19}

 \volnopage{ {\bf 2025} Vol., {\bf X} No. {\bf XX}, 000--000}
 \setcounter{page}{1}
 
\author{Hengkai Ding\inst{1,2}, Yiping Shu\inst{3}\thanks{E-mail: yiping.shu@pmo.ac.cn}, Yun Chen\inst{1,2}\thanks{E-mail: chenyun@bao.ac.cn}, Nan Li\inst{1,2},  Xiaoyue Cao\inst{2,1}, James Nightingale\inst{4,5,6}, Qiuhan He\inst{5}, Lei Wang\inst{3}, Hui Li\inst{7}}
  \institute{
   $^1$ National Astronomical Observatories, Chinese Academy of Sciences, 20A Datun Road, Chaoyang District, Beijing 100101, China;\\
   $^2$ School of Astronomy and Space Science, University of Chinese Academy of Sciences, Beijing 100049, China;\\
   $^3$ Purple Mountain Observatory, Chinese Academy of Sciences, No. 10 Yuan Hua Road, Nanjing 210023, China;\\
   $^4$ School of Mathematics, Statistics and Physics, Newcastle University, Newcastle upon Tyne, NE1 7RU, UK;\\
   $^5$ Institute for Computational Cosmology, Department of Physics, Durham University, South Road, Durham DH1 3LE, UK;\\
   $^6$ Centre for Extragalactic Astronomy, Department of Physics, Durham University, South Road, Durham DH1 3LE, UK;\\
   $^7$ Institute for Astrophysics, School of Physics, Zhengzhou University, Zhengzhou, 450001, China.\\
  } 
   
\vs \no
 {\small Received 2025 April 8; accepted 2025 April 21}


\abstract{
We present the first high-precision model for the group-scale strong lensing system CASSOWARY 19 (CSWA19), utilising images from the Hubble Space Telescope (HST). Sixteen member galaxies identified via the red-sequence method, and the main halo, all modelled as the dual Pseudo Isothermal Elliptical profile (dPIE), are incorporated into a parametric lens model alongside an external shear field. To model the system, we adopt the \textsc{PyAutoLens} software package, employing a progressive search chain strategy for realizing the transition of source model from multiple S\'ersic profiles to a brightness-adaptive pixelization, which uses 1000 pixels in the source plane to reconstruct the background source corresponding to 177,144 image pixels in the image plane. 
Our results indicate that the total mass within the Einstein radius is $M_{\theta_\mathrm{E}}$ $\approx 1.41\times10^{13}$~M$_{\odot}$ and the average slope of the total mass density $\rho (r)\propto r^{-\gamma}$ is $\tilde{\gamma}=1.33$ within the effective radius. This slope is shallower than those measured in galaxies and groups but is closer to those of galaxy clusters. 
In addition, our approach successfully resolves the two merging galaxies in the background source and yields a total magnification of $\mu=103.18^{+0.23}_{-0.19}$, which is significantly higher than the outcomes from previous studies of CSWA19.
In summary, our research demonstrates the effectiveness of the brightness-adaptive pixelization source reconstruction technique for modelling group-scale strong lensing systems. It can serve as a technical reference for future investigations into pixel-level modelling of the group- and cluster-scale strong lensing systems.
}

 \authorrunning{Hengkai Ding et al. }    
 \titlerunning{CSWA19 Modelling}  
 \maketitle

\section{Introduction}
\label{sec:introduction}
Strong gravitational lensing (hereafter strong lensing) occurs when a massive foreground object bends spacetime and deflects light rays, producing multiple images or distorted arcs of background sources. This phenomenon was first predicted by \citet{einstein1936lenslike}.
Strong lensing systems are typically categorized by mass and scale into three types: galaxy-scale, group-scale, and cluster-scale. Quantitatively, the classification is based on the halo mass $M_{200}$. For instance, systems with $M_{200}$ greater than $10^{14} M_{\odot}$ are classified as cluster-scale lenses, while those with $M_{200}$ less than or equal to $10^{13} M_{\odot}$ are considered galaxy-scale lenses. Group-scale lenses occupy the intermediate range. The boundaries between these categories—particularly between group- and cluster-scale lenses— are not strictly defined. In general, systems with a large number of member galaxies tend to have higher $M_{200}$. A strong lensing system is typically considered cluster-scale if it contains more than 25 member galaxies \citep{wang2024manga}. In this study, we define systems with more than one but fewer than 25 member galaxies as group-scale lenses, based on the complexity of lens modelling. It is important to note, however, that having fewer than 25 member galaxies does not necessarily imply $M_{200}<10^{14} M_{\odot}$. 

Strong lensing serves as a unique astrophysical laboratory with a wide range of scientific applications. 
One key application is measuring time delays between multiple images of strongly lensed quasars or supernovae, which provides a powerful probe to constraining the Hubble constant \citep{refsdal1964possibility, schechter1997quadruple, fassnacht2002determination, fohlmeister2007time, suyu2010dissecting, kelly2015multiple, millon2020cosmograil, wong2020h0licow, millon2020cosmograil, shajib2023tdcosmo, pascale2024sn}.
Besides, the statistics of strong lensing can be used in the cosmological analyses to estimate the cosmic curvature and other cosmological parameters \citep{Turner1984ApJ, Biesiada2006PhRvD,Cao2012AA,cao2012constraints,Chen2019MNRAS, Wang2020ApJ,Wei2022ApJ,Li2023PDU}.
As a natural cosmic telescope, strong lensing magnifies background sources, enabling the detailed study of distant galaxy structures \citep{yuan2011metallicity,wuyts2014magnified,jones2015grism,johnson2017stara,johnson2017star, cava2018naturea,dunham2019lens,mestric2022exploring,sharon2022cosmic,stacey2025nuclear}.
In addition, strong lensing is highly sensitive to the mass distribution of lensing objects, offering an independent and robust probe of both dark and luminous matter \citep{dalal2002directa, vegetti2009bayesian, vegetti2010quantifying, suyu2010halos, hezaveh2016detection, schuldt2019innera, gilman2020warm, meneghetti2020excessa, Du2020ApJ, wang2022constraining, Du2023ApJ, wang2024stronglensing, Lange2025}. 
For example, the Sloan Lens ACS Survey (SLACS) has modelled a large sample of galaxy-scale lenses and found that their average total density slope $\gamma$ is close to 2 \citep{auger2010sloan}, whereas cluster-scale systems typically exhibit shallower profiles \citep{newman2013density}.
\citet{newman2015luminous} extended this picture by incorporating group-scale lensing systems, suggesting that the variation in total density slopes reflects scale-dependent interactions between dark matter and baryons.
Group-scale strong lensing systems thus represent a crucial intermediate regime between galaxy- and cluster-scale lenses. Studying their dark matter distributions and galaxy properties provides valuable insight into the physical processes governing structure formation across cosmic scales.

Several surveys have targeted group-scale strong lensing systems, including the Sloan Bright Arcs Survey \citep{diehl2009sloan, kubo2009sloan}, the CFHTLS-Strong Lensing Legacy Survey (SL2S) - Arcs \citep{limousin2009newa, more2012cfhtlsstrong}, and the CAmbridge Sloan Survey of Wide ARcs in the Sky (CASSOWARY) \citep{belokurov2009two, stark2013cassowary}. Among them, CASSOWARY is a catalogue of strong lensing arcs identified from the Sloan Digital Sky Survey (SDSS), with a typical lens redshift of $\sim$0.4.
High-precision modelling is essential for detailed investigations of group-scale lenses. However, owing to the complex configurations often found in these systems, a standardized framework for their modelling is currently lacking. In this work, we introduce a practical modelling framework specifically designed for group-scale strong lensing systems.
In galaxy-scale lens modelling, extended image pixels have long been used to constrain parametric source model \citep{auger2009sloana,shu2016boss,shu2017sloan,cao2025csst}. With recent advances in modelling techniques, non-parametric pixelated source reconstructions have become widely adopted in the field of galaxy-scale strong lensing. These models can fit lensed image features down to the noise level and leverage thousands of pixels along the arcs to tightly constrain the lens mass distribution \citep{warren2003semilinear, suyu2006bayesian, nightingale2015adaptive, nightingale2024scanning, He2024}. 
Historically, modelling of group- or cluster-scale strong lenses typically relied solely on the positions of multiple lensed images to constrain the lens model. This approach was primarily adopted due to computational resource limitations. As a result, the wealth of information contained within the surface brightness distribution of the lensed images was not fully exploited (although recent studies have started to incorporate this detailed information, see e.g. \citealt{xie2024curling, acebron2024next, urcelay2024compact}).

In this work, we select CASSOWARY 19 (hereafter CSWA19) as a case study to demonstrate the pixel-level modelling of a group-scale strong lensing system. 
CSWA19, with Sloan ID SDSS J090002.79+223403.6, was first identified by \citet{diehl2009sloana}, who measured the source redshift and modelled the lens mass using a singular isothermal sphere (SIS), obtaining an Einstein radius of $7\farcs0 \pm 0.8$ and an enclosed mass of $(11.6\pm 2.7) \times 10^{12}~M_{\odot} $. \citet{stark2013cassowary} later adopted the same SIS model and reported a magnification of $\mu =6.5$. Subsequently, \citet{leethochawalit2016keck} employed the Light-Trace-Mass (LTM) method \citep{zitrin2015hubble}, deriving a lower magnification of $\mu=4.3$. Their mass model enabled source reconstruction from Keck/OSIRIS near-infrared spectroscopy, resolving the source plane velocity field and $H_{\alpha}$ distribution, which revealed an early-stage merging pair. Although follow-up spectroscopic observations and analysis of the source have been conducted \citep{rigby2018magellan, jones2018dust, mainali2023spectroscopy}, a detailed lens modelling study remains lacking.
In our work, we analyse high-resolution \textit{HST} imaging data of CSWA19, incorporating 177,144 image pixels into the $\chi^2$ statistic. This represents a major advancement over previous models of CSWA19, which relied solely on a few multiple image positions \citep{leethochawalit2016keck}. Compared to other group-/cluster-scale lensing studies that adopted pixel-based modelling, our analysis also involves a significantly larger number of pixels than \citet{urcelay2024compact} ($\sim$1500 pixels) and \citet{acebron2024next} ($\sim$78,000 pixels).

The structure of the paper is organized as follows. Section \ref{sec:data} presents the observational data of CSWA19, including optical/infrared imaging data and spectroscopy. Section \ref{sec:lens_modelling} outlines our modelling procedure using \textsc{PyAutoLens}, covering the selection of group members, modelling of the foreground galaxy light, the lens mass distribution, the source reconstruction, and the configuration of the search chain. Section \ref{sec:result} and Section \ref{sec:discussion} present our modelling results and discuss both the mass distribution of the lens and the intrinsic luminosity of the background source. Finally, Section \ref{sec:conclusion} summarizes our findings and discusses future directions.

Throughout this work, we assume a \textit{Planck} 2015 cosmological model \citep{planckcollaboration2016planck}. In such a cosmology, $1 \arcsec$ at redshift 0.48841 corresponds to 6.2099 kpc. All figures of this galaxy group are aligned to the WCS coordinate system: north is up, east is left. The reference centre in our analysis is fixed at the centre of the brightest group galaxy (hereafter BGG): R.A. = $9^{\text{h}}00^{\text{m}}02.79^{\text{s}}$, DEC = $+22^{\circ} 34'03.60''$. Magnitudes are given in the AB system.
  
\section{Data}
\label{sec:data}

    \subsection{Imaging data}
    \label{subsec:imaging_data}
    This study makes use of archival \textit{Hubble Space Telescope} (HST) high-resolution, multi-colour imaging (GO-11602; P.I.: Sahar Allam), from the \textsc{Wide Field Camera 3} (WFC3). The optical images were taken using WFC3/UVIS with filters F475W, F606W, and F814W, with a total exposure time of 5460, 2412 and 5598 seconds respectively in March 2010. The near-infrared images were taken by WFC3/IR with filters F110W and F160W, with two exposures in each filter and a total exposure time of 2412 and 2812 seconds respectively in April 2010. The UVIS field of view is approximately $2.8{'} \times 3{'}$, while that of the IR images is approximately $2.3{'} \times 2.1{'}$. Details of the data reduction process are provided in Section~\ref{subsec:data_preparation}.
    
    \subsection{Spectroscopic data}
    \label{subsec:spectroscopic_data}
    Three spectroscopic measurements are available for four member galaxies within a radius of $\sim14{''}$ from the reference centre, based on the \textit{Sloan Digital Sky Survey} (SDSS) DR16. As shown in Fig.\ref{fig:field_image}, two of these galaxies, the BGG and L6, were observed by the BOSS spectroscopic survey, which uses a $2{''}$-diameter fibre and provides spectral coverage from 3600$\sim$10400~\AA, with a resolution of 1560$\sim$2270 in the blue channel and 1850$\sim$2650 in the red channel. The third spectrum corresponds to a blended source identified as L1 and L2, observed by the SDSS Legacy survey using a $3{''}$ fibre with a wavelength range of 3800$\sim$9200~\AA~and resolution of 1850$\sim$2200. While treated as a single object in SDSS, L1 and L2 are clearly resolved in the HST imaging (see Fig.\ref{fig:field_image}).
    
    The spectroscopic redshifts of the BGG, L6 and L1/L2 are 0.48841$\pm$0.00012, 0.48367$\pm$0.00013 and 0.48912$\pm$0.00014, respectively. Velocity dispersion and stellar mass are retrieved from the SDSS database and are listed in Table~\ref{tab:F160W_members_properties}. For the purpose of lens modelling, we assume a common redshift of z=0.48841 for all member galaxies.
    
    For the background source, \citet{diehl2009sloana} obtained follow-up spectroscopic observations with the Astrophysical Research Consortium (ARC) 3.5~m telescope at the Apache Point Observatory, which shows emission and absorption lines. In this study, we adopt the average z=2.0325$\pm$0.0003 of the two knots(corresponding to the A and B images in Fig. \ref{fig:field_image}) as the source redshift.
    
    \subsection{Data preparation}
    \label{subsec:data_preparation}
    The science image data obtained directly from the HST archive consist of default multi-extension FITS products processed with the \textsc{Drizzlepac/AstroDrizzle} \citep{astrodrizzle} software to correct geometric distortions. For the UVIS images, we use the CTE-corrected, calibrated science data. The UVIS and IR science images obtained from the database have pixel scales of 0\farcs03962 and 0\farcs12825, respectively. We use the \textsc{reproject} package \citep{reproject} to resample the images with flux conservation and aligned them to a common WCS coordinate system for subsequent reduction and modelling. The resampled images have a pixel scale of 0\farcs04 for the UVIS images and 0\farcs128 for the IR images.
    
\begin{figure}
      \centering
      \includegraphics[width=0.8\textwidth]
      {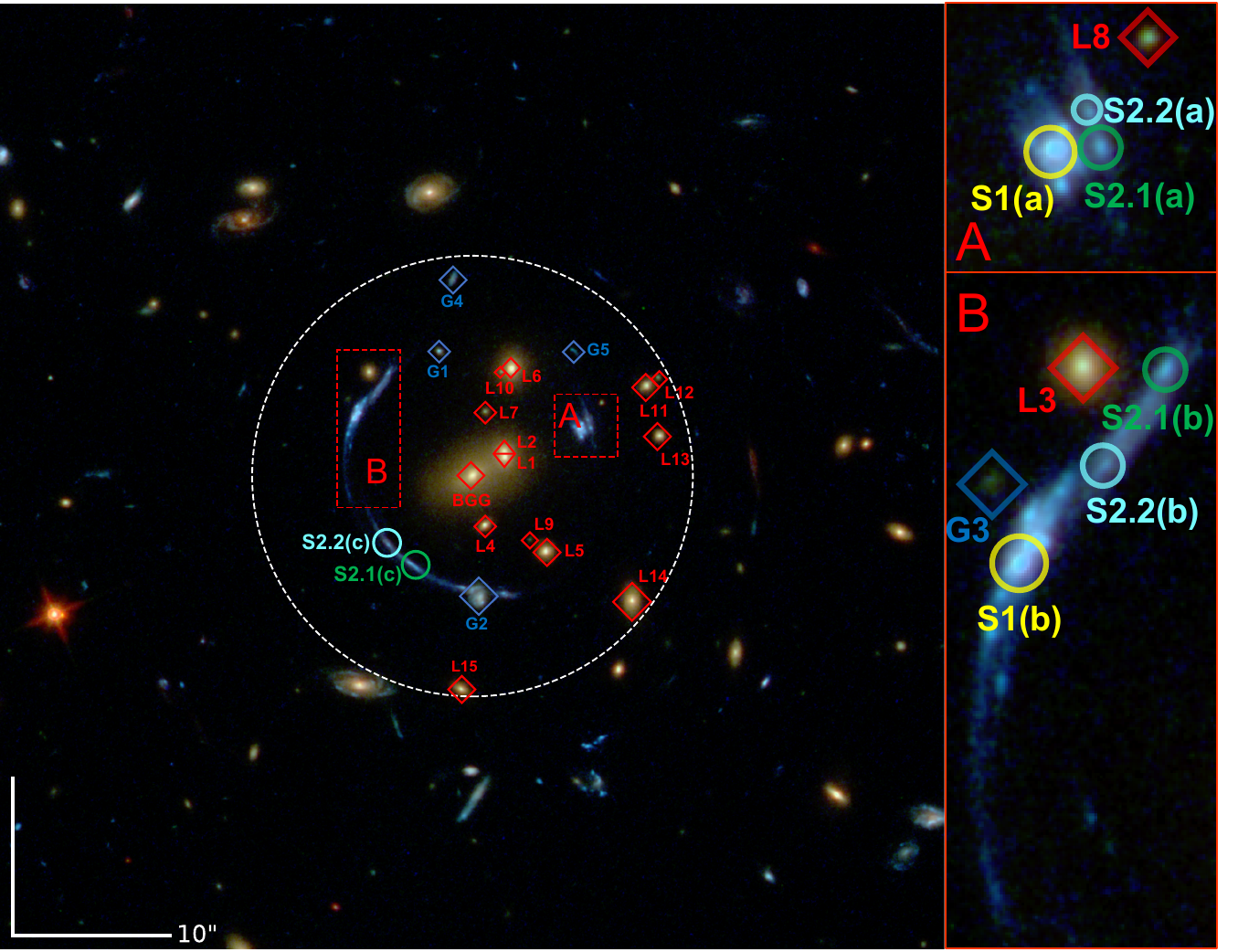} 
      \caption{Colour-composite image of the field of CASSOWARY 19 and its surrounding environment($1 \times 1 \text{ arcmin}^2$).
         The image is a combination of the \textit{F475W} (blue), \textit{F606W+F814W} (green), and \textit{F110W+F160W} (red) images. 
         We only consider member galaxies within a radius of $\sim 14{''}$ from the centre (marked with a white dashed circle). 
         The two right panels show the two images formed by the background source. 
         The lens galaxies are marked with red diamonds, and non-member galaxies within the circle are marked with blue diamonds. 
         The yellow circles mark the peak positions of the lensed images of the source galaxy 1. Accordingly, the green circles mark the peak positions of the lensed images of the source galaxy 2, and the cyan circles mark the second peak positions of the lensed images of the source galaxy 2.
      }
      \label{fig:field_image}
    \end{figure}

    After calculating and subtracting the uniform background using the \textsc{Photutils} package \citep{photutils_1_10_0}, we obtain the final images used for modelling. Based on the reduced multi-band images, we crop and generate a $1 \times 1 \text{ arcmin}^2$ colour-composite image centred on the reference centre (as described in Section \ref{sec:introduction}), shown in Fig.~\ref{fig:field_image}.
    
    The noise map is calculated following equations (12) and (13) of \citet{schuldt2019inner}, accounting for both background noise and Poisson noise:
    \begin{equation}
      \sigma_{\text{tot,i}}^2 = \sigma_{\text{bkg,i}}^2 + \sigma_{\text{poisson,i}}^2 = \sigma_{\text{bkg,i}}^2 + \left(\frac{\sqrt{d_i t_i}}{t_i} \right)^2
    \end{equation}
    where $t_i$ is the exposure time, $d_i$ is the intensity of pixel $i$ in units of electrons per second, and $\sigma_{\text{bkg,i}}$ is the constant background noise calculated from the empty area of the image.
    
    As noted by \citet{acebron2024next}, different from traditional cluster-scale lens modelling that only relies on multiple images positions, constructing an accurate point spread function (PSF) is critical for modelling extended lensed features. Here, we use the \textsc{psfr} \citep{psfr} software package to construct the PSF. Five isolated stars with diffraction spikes and minimal contamination are visually selected and cross-matched with the GAIA catalogue. Only one of them lies within the infrared field of view. For each star, we crop the image and corresponding noise map centred on the brightest pixel and use these as PSF samples. These are then stacked using \textsc{psfr} to generate the final PSF model. With the consideration of the calculating efficiency, the PSF is cropped to $31\times31$ pixels for UVIS and $35\times35$ pixels for IR.
    
    Considering both resolution and signal-to-noise ratio, we adopt the F475W-band image as constraint for lens modelling. This filter enhances the contrast of the bluer background source while minimizing contamination from the redder lens galaxies (see Fig. \ref{fig:field_image}). Conversely, the F160W-band image is employed to model the extended infrared light of the lens galaxies for the scaling relation described in Section \ref{sec:lens_modelling}. To accommodate this extended emission, the IR PSF is cropped to a larger size than that used for UVIS, despite the IR's coarser pixel scale.
    
\section{Lens modelling}
\label{sec:lens_modelling}
In our work, we use the \textsc{PyAutoLens} software \citep{pyautolens, autolens} to model the surface brightness distribution of the lens galaxies and source galaxies, as well as the mass distribution of the galaxy group. \textsc{PyAutoLens} is an open-source automated lens modelling tool based on the probabilistic programming software \textsc{PyAutoFit} \citep{pyautofit}, which is currently mainly applied to galaxy-scale strong gravitational lensing, with brightness-adapted pixel-grid-based semilinear inversion of background source reconstruction \citep{nightingale2015adaptive}. We apply \textsc{PyAutoLens}'s pixelized source reconstruction to model a group-scale gravitational lens system—its first application to a system with multiple lens galaxies—thereby extending the method's reach to more complex, larger-scale lenses and demonstrating its effectiveness in this regime.

Section \ref{subsec:membership_selection} presents the selection of member galaxies within the galaxy group using the red-sequence method \citep{gladders1998slope}. Section \ref{subsec:foreground_galaxy_light} shows the parametric modelling of the surface brightness distribution of the foreground galaxies, the subtraction of their light contamination, and the measurement of the lens galaxies’ luminosities in the F160W band. Section \ref{subsec:lens_mass_parametrization} describes the parametric modelling of the mass distribution.
Section \ref{subsec:source_model} subsequently details the source reconstruction techniques. Finally, Section \ref{subsec:sampling_and_search_chain} outlines our search strategy and sampling configuration.

    \subsection{Membership selection}
    \label{subsec:membership_selection}
    \begin{figure}
      \centering
      \includegraphics[width=0.8\textwidth]
      {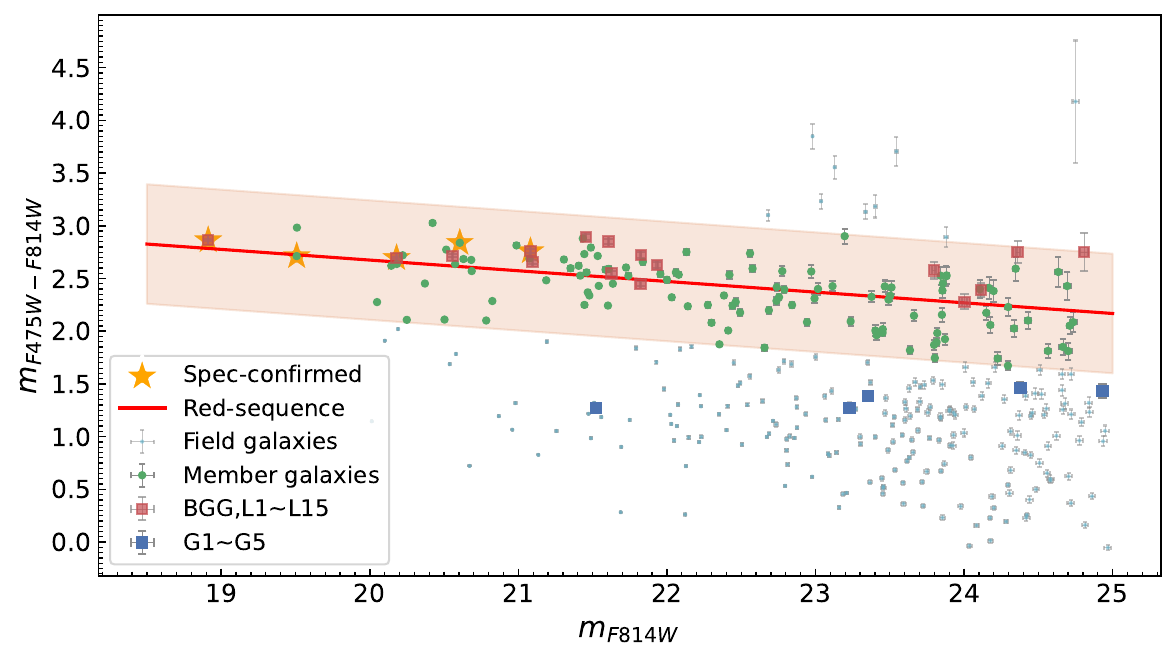} 
      \caption{Red-sequence fitting of the galaxy group. 
         The red solid line is the final converged red sequence, and the shaded area is the 3$\sigma$ range. 
         The 21 candidate galaxies are divided into two parts: member galaxies, i.e., BGG and L1$\sim$L15 (marked with red squares), and non-member galaxies, i.e.,G1$\sim$G5 (marked with blue squares).
         The yellow pentagrams denote the five member galaxies with spectroscopic redshifts used for the initial guess.
      }
      \label{fig:red_sequence}
    \end{figure}
    Before performing lens modelling and foreground galaxy light fitting, we first determine which galaxies are member galaxies of the group. \citet{diehl2009sloana} derived an Einstein radius of 7\farcs0 $\pm$ 0.8 for the system using a simplified SIS model. To balance computational efficiency and modelling accuracy, we only consider galaxies within $\sim 14{''}$ of the BGG's centre.  As shown in Fig. \ref{fig:field_image}, we identify a total of 21 candidate galaxies in this region using \textsc{Source Extractor} \citep{sextractor}, including the BGG.
    
    As in most group-scale lenses, not all members have spectroscopic confirmation. In this area, only BGG, L1, and L6 have spectroscopic redshift data (see Section \ref{subsec:spectroscopic_data}). For the remaining galaxies, we apply the so-called red-sequence method as our criterion to identify member galaxies.
    
    The red-sequence is a well-known colour-magnitude relation for member galaxies in galaxy clusters, where most early-type galaxies show a tight relation between colour and magnitude. This colour-magnitude relation was first noted by \citet{devaucouleurs1961integrated}, and latter systematically summarized by \citet{bower1992precision}. \citet{gladders1998slope} developed an algorithm to fit the red-sequence slope and demonstrated its utility in detecting galaxy clusters \citep{gladders2000new}.
    
    Following \citet{stott2009evolution} and \citet{gladders1998slope}, we adopt their slope-fitting method, using F475W-F814W as our colour index since these bands bracket the 4000 \AA{} break at the group's redshift, effectively distinguishing red-sequence galaxies from star-forming galaxies \citep{stott2009evolution}. We use \textsc{Source Extractor} to measure Kron magnitudes and errors for all galaxies in the $2.8{'} \times 3{'}$ field.
    
    We construct the red sequence through the standard iterative procedure of \textit{linear fitting, slope correction, single Gaussian fitting, and 3-sigma clipping} \citep{stott2009evolution,gladders1998slope}.
    The final red sequence is shown in Fig. \ref{fig:red_sequence} as a red solid line, with the 3$\sigma$ range shaded. This classification divides the 21 candidate galaxies into members (BGG and L1$\sim$L15, marked with red squares) and non-members (G1$\sim$G5, marked with blue squares). Their positions and labels are indicated in Fig. \ref{fig:field_image}. All member galaxies have F160W magnitudes brighter than 24 mag. In the subsequent modelling, we only consider member galaxies BGG and L1$\sim$L15.
    
    \subsection{Foreground galaxy light}
    \label{subsec:foreground_galaxy_light}
    \begin{figure}
      \centering
      \begin{subfigure}[b]{1.0\textwidth}
         \centering
         \includegraphics[width=\textwidth]{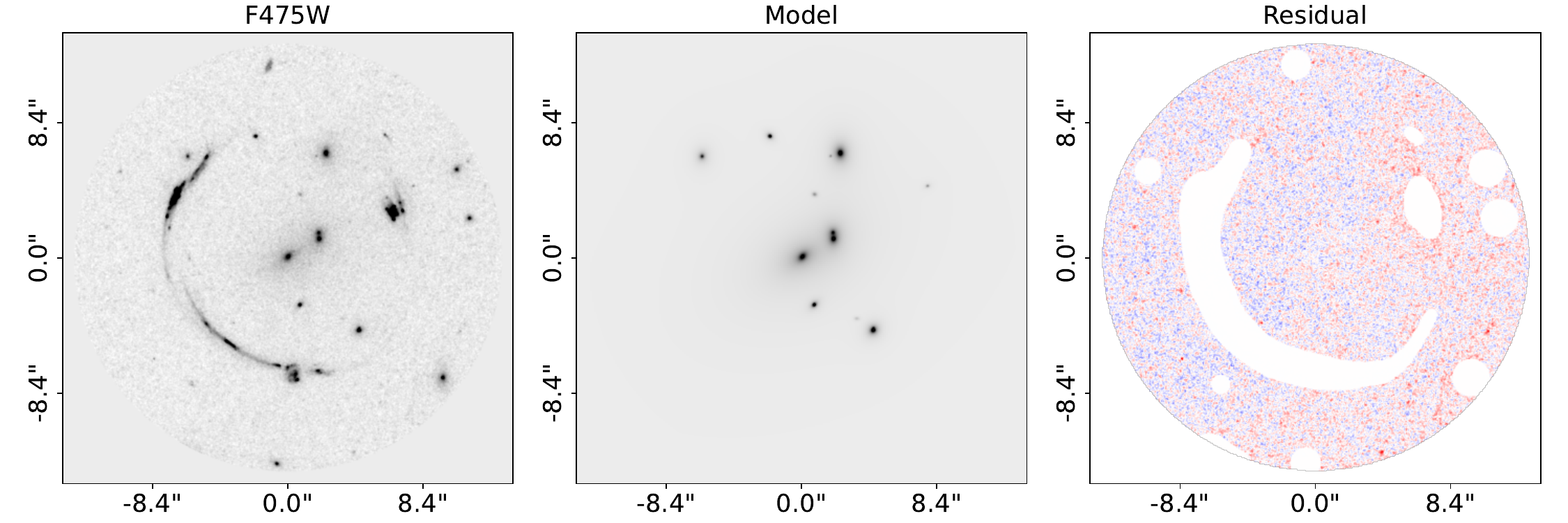}
         \label{fig:3.1}
      \end{subfigure}
      \vspace{1em}
      \begin{subfigure}[b]{1.0\textwidth}
         \centering
         \includegraphics[width=\textwidth]{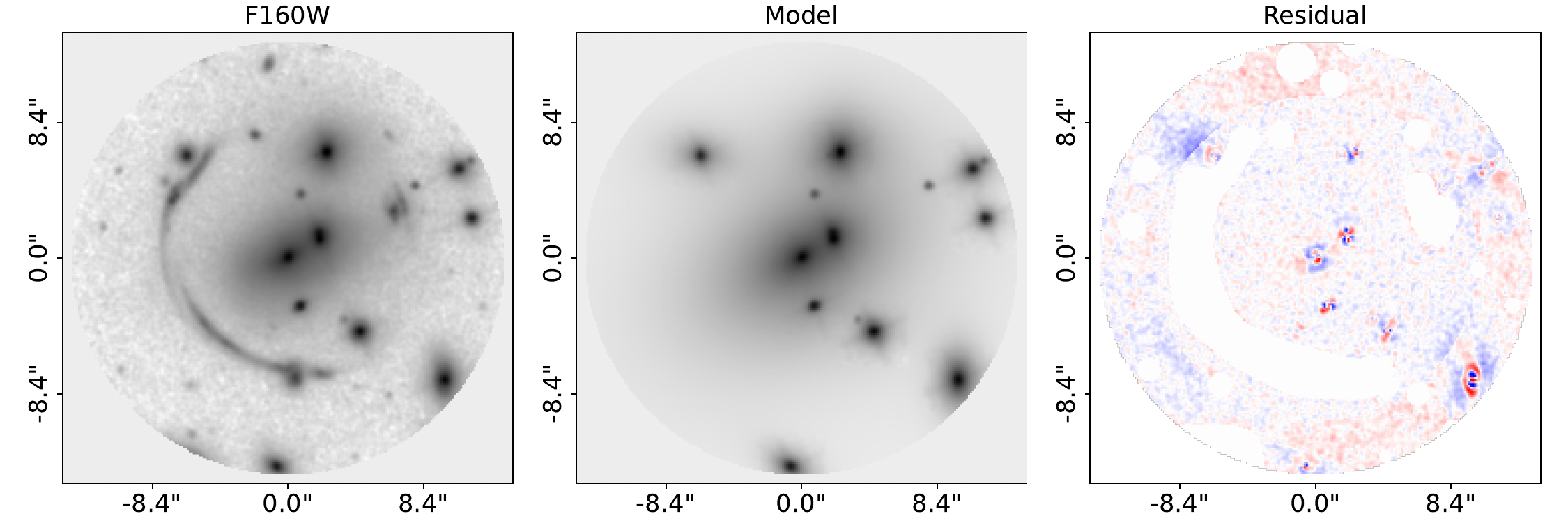}
         \label{fig:3.2}
      \end{subfigure}
      \caption{
         Multi-S\'ersic light profile fittings to the F475W and F160W images for the lens galaxies.
         The upper panels show the F475W band fitting results, where the panels from left to right show the observed image, the S\'ersic model image, and the normalized residual image (range $\pm$5$\sigma$). 
         The lower panels show the F160W band fitting results, and the panels from left to right are the the observed image, S\'ersic model image, and the normalized residual image (range $\pm$10$\sigma$). 
         Only the BGG, L1$\sim$L10, and G1 are modelled with the F475W band image, because the purpose of the F475W band modelling is to subtract the light contamination of the foreground galaxies, and the light contamination of L11$\sim$L15 and G2$\sim$G5 is small or outside the lens modelling image area.
         }
      \label{fig:combined_lens_light}
    \end{figure}
    
    \begin{table}
    \caption{Position, magnitude, effective radius fitting results with 1-$\sigma$ uncertainty for the member galaxies in the F160W band, and the dispersion, stellar mass measurement from SDSS}
    \label{tab:F160W_members_properties}
    \centering
    \begin{threeparttable}
    \setlength{\tabcolsep}{1.5mm}
    \begin{tabular}{c r r c c c c c c} 
    \hline
    \multirow{2}{*}{ID} & \multirow{2}{*}{$x$ [${''}$]} & \multirow{2}{*}{$y$ [${''}$]} & \multicolumn{2}{c}{$R_\mathrm{eff}$[$\mathrm{kpc}$]} & \multicolumn{2}{c}{AB-mag} & \multirow{2}{*}{$\sigma$[$\mathrm{km/s}$]} & \multirow{2}{*}{$\log(M_*/M_\odot)$} \\ \cline{4-7}
                              &                               &                               & Best-fit & Marginalized & Best-fit & Marginalized &                  &                       \\ 
    \hline
    BGG                      & $0.0236$                     & $-0.0087$                    & $21.707$ & $21.704^{+0.068}_{-0.056}$ & $17.126$ & $17.126^{+0.002}_{-0.002}$ & $263^{+44}_{-44}$ & $12.01^{+0.09}_{-0.07}$ \\
    L1                       & $1.2015$                     & $2.0348$                     & $3.470$  & $3.601^{+0.065}_{-0.068}$  & $18.771$ & $18.748^{+0.011}_{-0.011}$ & $157^{+33}_{-33}$ & $11.56^{+0.09}_{-0.09}$ \\
    L2                       & $1.6308$                     & $1.9956$                     & $2.368$  & $3.217^{+0.080}_{-0.090}$  & $19.728$ & $19.758^{+0.016}_{-0.016}$ & --                 & --                         \\
    L3                       & $6.5766$                     & $-6.5502$                    & $7.141$  & $7.185^{+0.110}_{-0.098}$  & $19.660$ & $19.658^{+0.007}_{-0.007}$ & --                 & --                         \\
    L4                       & $-3.1112$                    & $0.7814$                     & $0.392$  & $0.378^{+0.010}_{-0.011}$  & $19.566$ & $19.576^{+0.013}_{-0.013}$ & --                 & --                         \\
    L5                       & $-4.7812$                    & $4.6346$                     & $1.674$  & $1.693^{+0.013}_{-0.012}$  & $19.452$ & $19.452^{+0.006}_{-0.005}$ & --                 & --                         \\
    L6                       & $6.7970$                     & $2.4680$                     & $5.211$  & $5.176^{+0.024}_{-0.020}$  & $18.468$ & $18.470^{+0.002}_{-0.002}$ & $155^{+32}_{-32}$ & $11.61^{+0.10}_{-0.09}$ \\
    L7                       & $4.0817$                     & $0.8063$                     & $0.639$  & $0.588^{+0.049}_{-0.070}$  & $22.686$ & $22.636^{+0.042}_{-0.081}$ & --                 & --                         \\
    L8                       & $4.6237$                     & $8.1924$                     & $0.121$  & $0.197^{+0.125}_{-0.074}$  & $21.381$ & $21.885^{+0.616}_{-0.548}$ & --                 & --                         \\
    L9                       & $-4.0195$                    & $3.6123$                     & $0.423$  & $0.307^{+0.143}_{-0.151}$  & $23.639$ & $23.909^{+0.293}_{-0.767}$ & --                 & --                         \\
    L10                      & $6.6180$                     & $1.7786$                     & $0.210$  & $0.174^{+0.074}_{-0.049}$  & $22.467$ & $22.828^{+0.649}_{-0.659}$ & --                 & --                         \\
    L11                      & $5.7239$                     & $11.0402$                    & $1.526$  & $1.534^{+0.023}_{-0.023}$  & $20.150$ & $20.150^{+0.013}_{-0.013}$ & --                 & --                         \\
    L12                      & $6.2343$                     & $11.7888$                    & $0.352$  & $0.337^{+0.159}_{-0.156}$  & $22.888$ & $22.916^{+0.357}_{-0.604}$ & --                 & --                         \\
    L13                      & $2.5430$                     & $11.8803$                    & $0.418$  & $0.425^{+0.010}_{-0.009}$  & $20.070$ & $20.079^{+0.012}_{-0.008}$ & --                 & --                         \\
    L14                      & $-7.9041$                    & $10.1086$                    & $2.886$  & $2.888^{+0.013}_{-0.013}$  & $18.940$ & $18.940^{+0.004}_{-0.004}$ & --                 & --                         \\
    L15                      & $-13.5111$                   & $-0.7457$                    & $1.927$  & $1.920^{+0.025}_{-0.024}$  & $19.996$ & $20.001^{+0.006}_{-0.004}$ & --                 & --                         \\
    \hline
    \end{tabular}
    \end{threeparttable}
    \begin{tablenotes}
      \item \textbf{Note}: The above position, magnitude, effective radius are all from the multi-S\'ersic fitting of the F160W band, and the velocity dispersion and stellar mass are from the SDSS measurement. Note that only three member galaxies have spectroscopic data, corresponding to BGG, L6, and L1/L2, L1 and L2 are identified as a single source in SDSS, but can be resolved in our high-resolution HST images. The stellar masses from SDSS database are using the FSPS model \citep{conroy2009propagation} with Granada method in \texttt{early-star-formation} version for dust extinction.
    \end{tablenotes}
    \end{table}
    
    All the light profiles of the foreground galaxies in this work are described by one or more elliptical S\'ersic profiles \citep{sersic1963influence}:
    \begin{equation}
      I(\xi) = I_0 \exp {\left\{- k \left[\left(\frac{ \xi}{R_e}\right)^{\frac{1}{n}} - 1\right] \right\}}
    \end{equation}
    where $\xi = \sqrt{q x^2 + y^2/q}$, and $I_0$ is the intensity at the effective radius $R_e$, $n$ is the S\'ersic index, and $k$ is a constant that depends on $n$ \citep{ciotti1999analytical}.
    The ellipticity is introduced as described in \citet{nightingale2024scanning} equation (1): 
    \begin{equation}\label{eq:ell_comps}
      e_1 = \frac{1 - q}{1 + q} \sin 2\phi, e_2 = \frac{1 - q}{1 + q} \cos 2\phi
    \end{equation}
    where $q$ is the axis ratio, and $\phi$ is the position angle defined counterclockwise from the positive x-axis. Therefore, a single S\'ersic profile has seven free parameters: $I_0$, $R_e$, $n$, $x$, $y$, $e_1$, and $e_2$. To reduce parameter space complexity, we use the \texttt{linear light profile} in \textsc{PyAutoLens} (unless stated otherwise), where intensities are solved via linear algebra during model fitting rather than treated as free parameters in the non-linear sampling (see \citealt{He2024} for details). This approach employs a non-negative least squares solver \citep{bro1997fast}, which enforces physically meaningful, positive intensity values.
    
    In modelling the light of foreground galaxies, we iteratively determine the number of S\'ersic components required to describe the light of each member galaxy. Each galaxy is initially modelled with a single S\'ersic component, and additional S\'ersic components are added one by one until the intensity of the newly added component is solved to be zero.
    Unless otherwise stated, the centres of all S\'ersic components associated with a given galaxy are fixed to the same position.
    
    For the F475W band fitting, our primary goal is to subtract the light contamination from foreground galaxies. Therefore, we only fit the galaxies located within or near the arc: BGG, L1 $\sim$ L10, and G1. Since the light distribution in the F475W band is relatively compact, most galaxies could be well modelled individually within local circular regions. An exception is made for the central galaxies BGG, L1, L2, and L4, whose light profiles overlap significantly and are thus fitted simultaneously.  The F475W band fitting results are reliable, with only BGG and L5 requiring two S\'ersic components, while the others are well described by a single component. The upper panel of Fig. \ref{fig:combined_lens_light} shows the F475W band fitting results, displaying, from left to right, the observed image data, the S\'ersic model image, and the normalized residual map.  The residual image masks the arc and galaxies outside the modelling region.
    
    For the F160W band, our goal is to obtain photometry for each lens galaxy, which is required for the scaling relation discussed in Section \ref{subsec:lens_mass_parametrization}. Owing to the more extended light distribution in this band, we divide the galaxies into three modelling batches. We first fit BGG and L1$\sim$L6 simultaneously, masking the arc and other galaxies, since the extended light of these inner galaxies contaminated both the arc and nearby faint or outer galaxies. 
    A simple uniform background subtraction is applied during the data reduction (see Section.~\ref{subsec:data_preparation}), but residual non-uniform background remains during lens light modelling. This component may be attributed to inhomogeneous intra-cluster light (ICL). Following the approach of \citet{martis2024modelling}, we iteratively estimate and subtract this background using \textsc{Photutils} within the modelling area. After removing the inner galaxy light and non-uniform background, the faint galaxies L7$\sim$L10 and the distant galaxies L11$\sim$L15 are fitted independently, in a manner consistent with the F475W band modelling.
    
    The lower panel of Fig. \ref{fig:combined_lens_light} shows the F160W band fitting results; from left to right are the observed image data, the S\'ersic model image, and the normalized residual map, with the arc and local unknown fluctuations masked in the residual image. The fitting results are not as reliable as the F475W band, with some residuals remaining in the galaxy cores due to the more complex light structures present in the F160W band. We calculate the residual flux in the central regions of these galaxies, and in all cases, the ratio of residual flux to the total luminosity of the corresponding galaxy is less than 1\%.  Table \ref{tab:F160W_members_properties} shows the S\'ersic fitting photometry results of the foreground galaxies in the F160W band, where the coordinates are taken as the maximum likelihood values in the reference coordinate system, and the magnitudes are reported as the maximum likelihood values along with the median and 1$\sigma$ error of the marginal distributions.
    
    \subsection{Lens mass parametrization}
    \label{subsec:lens_mass_parametrization}
    Regarding the lens mass model, the mass distribution of the galaxy group is assumed to be dominated by a group-scale dark matter halo centred on the Brightest Group Galaxy (BGG), with member galaxies residing within their own dark matter subhalos embedded in this main halo.
    For the above objects, we adopt a dual Pseudo Isothermal mass profile with pseudo-ellipticity\footnote{This profile was introduced into \textsc{PyAutoGalaxy} by O'Donnell and has also been applied to the modelling of the galaxy cluster MACS J0138 \citep{odonnell2025constraint}.}, implemented in a development branch\footnote{The corresponding implementation is available at \url{https://github.com/jhod0/PyAutoGalaxy/tree/dPIE}.} of \textsc{PyAutoGalaxy} \citep{nightingale2023pyautogalaxy}, the modelling sub-package of \textsc{PyAutoLens}. The functional form of the convergence $\kappa$ is: 
    \begin{equation}\label{eq:kappa_dpie_al}
      \kappa(\xi) = \frac{E_0}{2}\frac{r_s + r_a}{r_s} \left( \frac{1}{\sqrt{r_a^2 + \xi^2}} - \frac{1}{\sqrt{r_s^2 + \xi^2}} \right) (1 - \mathcal{A}) + \frac{\alpha(\xi)}{\xi} \mathcal{A}
    \end{equation}
    where $r_a$ is the core radius, $r_s$ is the scale radius. For the 3D density distribution, in the core region where $r < r_a$, it has a relatively flat mass distribution, in the transition region where $r_a < r < r_s$, it behaves like an Isothermal with $\rho \sim r^{-2}$, and in the outer region where $r > r_s$, it provides a drop-off with $\rho \sim r^{-4}$.
    The lens strength $E_0$ is defined as:
    \begin{equation}
      E_0 = 6\pi \frac{D_{LS}D_{L}}{D_{S}}\frac{\sigma_{\mathrm{dPIE}}^2}{c^2}
    \end{equation}
    As described by \citet{eliasdottir2007where}, this definition of $E_0$ is consistent with \textsc{Lenstool} \citep{kneib1996hubble,jullo2007bayesian}, where $E_0$ is proportional to $\sigma_{\mathrm{dPIE}}^2$, and has units of length.
    $\mathcal{A}$ is the asymmetric term with respect to $x$ and $y$, defined as:
    \begin{equation}
      \mathcal{A} = \frac{\epsilon(1-\epsilon)x^2 - \epsilon(1+\epsilon)y^2}{\xi^2}
    \end{equation}
    where $\xi$ is defined as:
    \begin{equation}
      \xi = \sqrt{x^2(1-\epsilon) + y^2(1+\epsilon)}
    \end{equation}
    The ellipticity $\epsilon$ is defined as: $\epsilon = \sqrt{e_1^2 + e_2^2} = (1-q)/(1+q)$ where $e_1$ and $e_2$ are defined as in Eq.~(\ref{eq:ell_comps}). $\alpha(\xi)$ is the deflection angle in the circular case at radius $\xi$, defined as in Eq.~(A19) of \citet{eliasdottir2007where}. 
    
    This profile introduces ellipticity to the projected lensing potential rather than to the projected surface mass density of the standard dual Pseudo Isothermal Elliptical (dPIE, \citet{eliasdottir2007where}) profile, an approximation often referred to as \texttt{pseudo-ellipticity} \citep{kovner1987marginal, golse2002pseudo}. This approximation allows for easier computation of the gradient and Hessian of the lens potential \citep{kovner1987marginal, golse2002pseudo}, although it is known that the surface density (Eq.~\ref{eq:kappa_dpie_al}) becomes dumbbell-shaped at higher ellipticities \citep{kassiola1993elliptica, golse2002pseudo, shajib2019unified}, which is unphysical. In our F160W-band modelling, the axis ratios of the member galaxies' light distributions are all greater than 0.6, justifying the use of this approximation. 
    
    In the circular case ($\epsilon = 0$, $\mathcal{A} = 0$), the above formulation reduces to the standard form of the dPIE convergence profile:
    \begin{equation}\label{eq:dPIE}
      \kappa(\xi) = \frac{E_0}{2}\frac{r_s + r_a}{r_s} \left( \frac{1}{\sqrt{r_a^2 + \xi^2}} - \frac{1}{\sqrt{r_s^2 + \xi^2}} \right)
    \end{equation}
    The dPIE profile is equivalent to the Pseudo-Jaffe model when $r_a = 0$ \citep{dalal2002direct, vegetti2010detection}, to the PIEMD model when $r_s \rightarrow \infty$ \citep{kassiola1993elliptic}, and to the SIE model when both $r_a = 0$ and $r_s \rightarrow \infty$, in which case $E_0$ represents the Einstein radius.
    
    Therefore, a complete description of the dPIE model with pseudo-ellipticity requires seven parameters: $E_0$, $r_a$, $r_s$, $x$, $y$, $e_1$, $e_2$. For each member galaxy, we use a dPIE model to describe their total mass distribution. To reduce the complexity of the parameter space, we follow \citet{grillo2015clashvlt}, \citet{chirivi2018macs}, and \citet{wang2022constraining}, assuming that all member galaxies follow the following scaling relation:
    \begin{equation}\label{eq:scaling_relation}
      \sigma_{\mathrm{dPIE},i}^2 = \sigma_{\mathrm{dPIE},\mathrm{ref}}^2\left( \frac{L_i}{L_{\mathrm{ref}}} \right)^{0.7}\quad \text{and} \quad r_{s,i} = r_{s,\mathrm{ref}}\left( \frac{L_i}{L_{\mathrm{ref}}} \right)^{0.5}
    \end{equation}
    This specific scaling relation assumes that the total mass-to-light ratio increases with luminosity \citep{grillo2015clashvlt}: $M_{\mathrm{total}}/L \propto L^{0.2}$, known as the tilt of the Fundamental Plane of early-type galaxies \citep{faber1987globala, bender1992dynamically}. Some works have shown that it can better reconstruct the observed positions of multiple images \citep{grillo2016story, kelly2016deja}. In Eq. (\ref{eq:scaling_relation}), $L_i$ is the F160W band luminosity of the $i$-th galaxy, and $L_{\mathrm{ref}}$ is the F160W band luminosity of the reference galaxy. The F160W band luminosity serves as a good proxy for their total mass \citep{grillo2015clashvlt}. We choose the second brightest galaxy L6 in the system as the reference galaxy instead of BGG, since BGG may have undergone different formation and evolution processes, thus not following the usual elliptical galaxy scaling relation \citep{caminha2019strong,richard2021atlas}.
    
    We also incorporate an external shear field to account for the tidal distortion induced by the surrounding environment or other mass components of CSWA19 beyond the modelled region. In \textsc{PyAutoLens}, this shear field is described by two elliptical components: $(\gamma_{\mathrm{ext},1}, \gamma_{\mathrm{ext},2})$. Their relationship with the shear magnitude $\gamma_{\mathrm{ext}}$ and the orientation measured counterclockwise from north $\phi_{\mathrm{ext}}$ is as follows:
    \begin{equation}
      \gamma_{\mathrm{ext}} = \sqrt{\gamma_{\mathrm{ext},1}^2 + \gamma_{\mathrm{ext},2}^2} \quad \text{and} \quad \phi_{\mathrm{ext}} = \frac{1}{2} \arctan \left( \frac{\gamma_{\mathrm{ext},2}}{\gamma_{\mathrm{ext},1}} \right)
    \end{equation}
    
    For the member galaxies, we assume a circularly symmetric dPIE model, and fix their mass and light centres to further reduce the number of free parameters. For the BGG and the main halo of the group scale, we use an elliptical dPIE model. Both the BGG and member galaxies adopt a vanishing core dPIE, which approximates a singular isothermal profile in both the core and transition regions. We also fix the mass centre at the inferred light centre of BGG. 
    The scale radius $r_s$ of the main halo is fixed to a large value of $1000{''}$, because this parameter describes the nature of the mass distribution of the main halo at large scales \citep{richard2021atlas,bergamini2021new}, which is usually much larger than the radius that a strong lens can provide constraints in practice \citep{limousin2007combining}.

    We now explain the choices of the adopted priors for our initial model (see Model~1 in Table~\ref{tab:search_chain}). First is the scale radius $r_s$ for BGG and member galaxies, for which we set an upper limit to consider the tidal stripping of the dark matter halo \citep{limousin2007truncation, limousin2009probing, natarajan2009survival, wetzel2010what}. Based on prior experience, we set this upper limit to approximately 5 times the half-light radius in the F160W band and adopt a broader prior range to allow flexibility. For the parameters of the main halo, we use uniform priors; specifically, we limit its core radius $r_a$ to $3{''}\sim17{''}$. This wide range is inspired by the results of \citet{richard2021atlas} and \citet{bergamini2021new} for extended halos, to exclude unrealistically large core radii.  
    The Gaussian prior on the velocity dispersion $\sigma_{\mathrm{dPIE}}^{\mathrm{BGG}}$ (also corresponding to $E_0^{\mathrm{BGG}}$) is informed by the spectroscopic measurement from SDSS (see Sec. \ref{subsec:spectroscopic_data}), with the prior mean set to the measured value and the standard deviation taken as three times the reported measurement uncertainty.
    Note that the theoretical velocity dispersion $\sigma_{\mathrm{dPIE}}$ defined by \citet{eliasdottir2007where} is not exactly equal to the observed velocity dispersion, but differs by a projection coefficient $c_p$, which depends on the core radius $r_a$, the scale radius $r_s$, and the aperture radius $R$ (see Fig. C.3 and Eq. C.16 of \citet{bergamini2019enhanced}).

    It is also worth noting that the member galaxy L1 is the third brightest and closest to BGG, which may deviate significantly from the scaling relation. We use the same elliptical dPIE model to describe its mass distribution as BGG, and allow its $r_s$ and $E_0$ to vary freely.
   
    \subsection{Source model}
    \label{subsec:source_model}
    \begin{figure}
      \centering
      \includegraphics[width=0.6\textwidth]
      {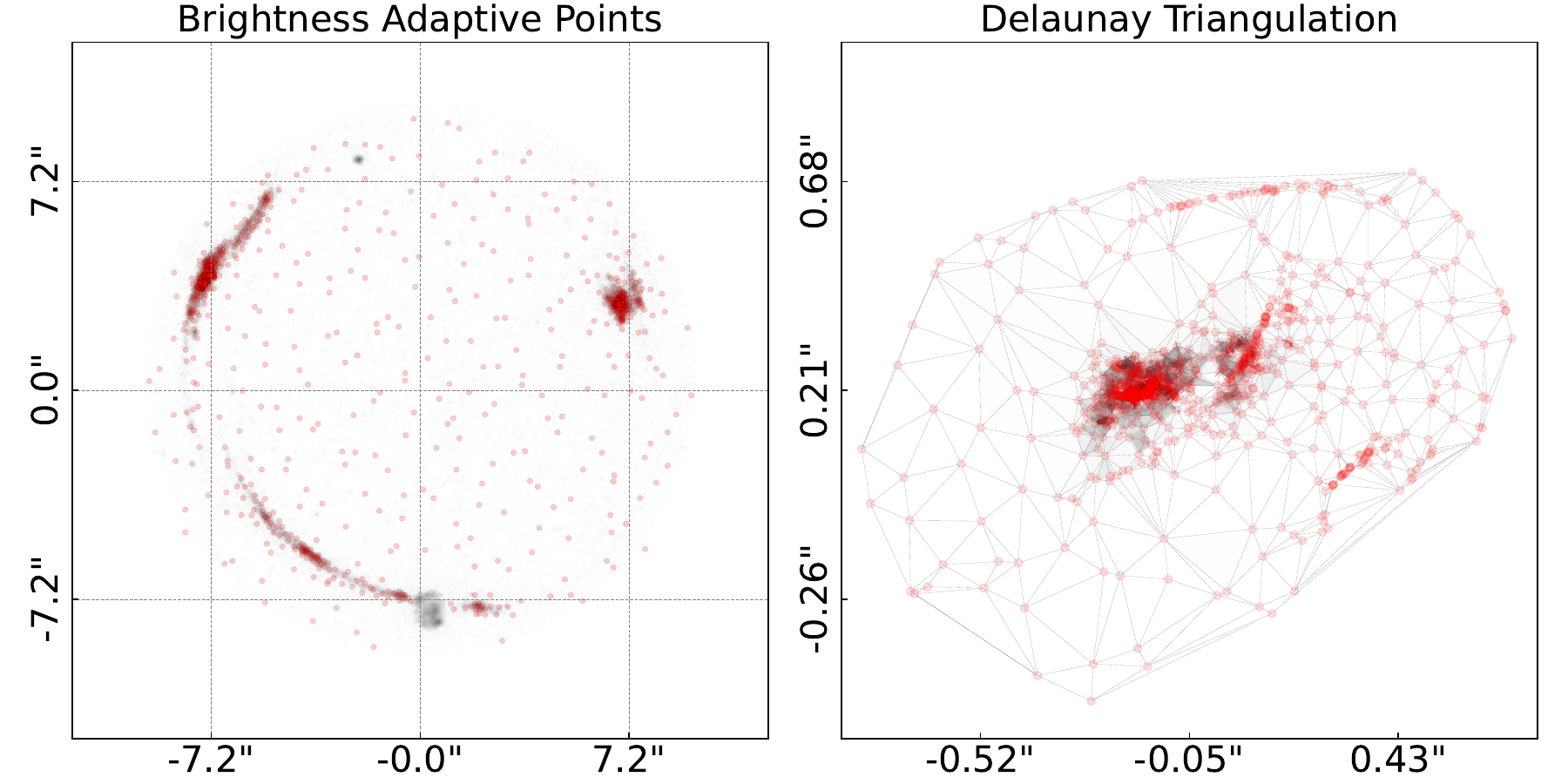} 
      \caption{Schematic diagram of the construction of the brightness-adaptive grid. The left image shows the irregular grid point distribution on the image plane, constructed by the Hilbert space-filling curve, these 1000 points are distributed more densely in brighter areas and sparsely in unstructured areas, this grid point distribution comes from the best fitting values of \textbf{Model~5-0}, and we manually mask the area of G2; the right image shows the Delaunay triangle mesh on the source plane, the vertices of these triangles are mapped from the points on the image plane to the source plane through the mass model, the grid points shown here correspond to the mass distribution of the best fitting results of \textbf{Model~5-1}.}
      \label{fig:adaptive_grid}
    \end{figure}

    We fit lens models using two different approaches for the source galaxy. The simpler model uses one or more S\'ersic profile–using the same elliptical formulation as described in Section \ref{subsec:foreground_galaxy_light}–to characterize the background galaxy's surface brightness. When multiple components are implemented, their centroids and ellipticities become free parameters in the fitting process.

    Advanced modelling employs an adaptive pixelization to reconstruct the irregular surface brightness distribution of the background source. A uniform grid of size $y_{\mathrm{pixels}} \times x_{\mathrm{pixels}}$ is overlaid on the image plane, from which $J = y_{\mathrm{pixels}} \times x_{\mathrm{pixels}}$ cell centres are identified as those within the image mask. These centres are mapped to the source plane using deflection angles from the lens model and form the vertices of a Delaunay triangle mesh for source reconstruction, following the method of \citet{vegetti2009bayesian}. This approach introduces only three non-linear parameters: the x and y dimensions of the image-plane mesh and the regularization coefficient $\lambda$. The mesh naturally adapts to the lensing magnification pattern \citep{nightingale2015adaptive}.

    However, adapting to magnification alone can lead to sub-optimal reconstructions if the source lies in low-magnification regions, away from the caustic. To address this, \textsc{PyAutoLens} includes a brightness-based pixelization scheme that instead adapts to the source’s surface brightness, using a model-predicted image of the lensed source or observed image itself with lens light subtracted called the ``adapt image''. A normalized weight map is computed from this image, assigning higher weights to brighter regions. A total of $I$ image-plane coordinates are then probabilistically sampled from this map using a generalized Hilbert space-filling curve \citep{hilbert1891stetige}.

    This scheme better captures the source’s structure (see Fig.~\ref{fig:adaptive_grid}), yielding more detailed reconstructions with fewer vertices and significantly reduced computational cost. It has recently outperformed the previously used KMeans clustering method in modelling Euclid-resolution images \citep{wang2025measuring}. A full description of this new \texttt{HilbertMesh} method is presented in \citet{he2025notb}. 
   
    For a given mass model and set of source pixel centres, the mapping matrix $f_{ij}$ is constructed. Assuming that there are $J$ image-pixels in the masked observed data, this matrix maps the $j$-th pixel on the image plane to the $i$-th triangle on the source plane one by one. By default, \textsc{PyAutoLens} uses a 4$\times$4 subgridding for each image plane pixel, meaning that the actual number of image plane sub-pixels used for mapping is 16$\times J$. In general, multiple image plane pixels are mapped to the same source plane triangle, and $f_{ij}$ records the weights of each image pixel-vertex pair mapping, which are calculated by triangle linear interpolation: 
    \begin{equation}
      \hat{z}_j = \frac{A_{j i_{2} i_{{3}}} z_{i_{1}} + A_{i_{1} j i_{{3}}} z_{i_{2}} + A_{i_{1} i_{{2}} j} z_{i_{3}}}{A_{i_{1} i_{2} i_{3}}}
    \end{equation}
    where the $j$-th sub-pixel falls into the triangle with vertices $i_{1}$, $i_{2}$, and $i_{3}$, $A$ is the area of the triangle formed by the three points, and $z$ is the value of the corresponding point. $f_{ij}$ is convolved with the PSF to obtain a blurred mapping matrix to include the effect of the PSF, and we assume that $f_{ij}$ from here on includes the PSF convolution.

    Therefore, assuming that the value of the source plane vertex is $s_i$ and the flattened observed data is $d_j$, we can obtain the best source reconstruction by minimizing the following equation:
    \begin{equation}
    \label{eq:Chi2_for_inversion}
      \chi^2 = \sum_{j=1}^{J} \left( \frac{d_j - \sum_{i=1}^{I} f_{ij}s_i}{\sigma_j} \right)^2
    \end{equation}
    where $\sigma_j$ is the error of the observed data, i.e., the noise map obtained in Section \ref{subsec:data_preparation}. We consider that the $d_j$ here has already been subtracted the light of the foreground galaxies.
    The minimization solution of the above equation can be abstracted into the following matrix form:
    \begin{equation}
    \label{eq:inversion}
      \mathbf{s} = \mathbf{F}^{-1} \mathbf{D}
    \end{equation}
    where $\mathbf{F}$ is the curvature matrix with $F_{ik} = \sum_{j=1}^J f_{ij}f_{kj}/\sigma^2_j$, and $\mathbf{D}$ is the data vector with $D_i = \sum_{j=1}^J d_j f_{ij}/\sigma^2_j$. Therefore, the source reconstruction can always be solved by linear algebra methods. Combined with the parametric mass model, this method is the so-called semilinear inversion (first proposed by \citet{warren2003semilinear}).
    To avoid overfitting, a regularization term $\lambda H$ is introduced on the basis of the above:
    \begin{equation}
    \label{eq:inversion_with_reg}
      \mathbf{s} = [\mathbf{F} + \lambda \mathbf{H}]^{-1} \mathbf{D}
    \end{equation}
    where $\mathbf{H}$ is the regularization matrix, and $\lambda$ is the regularization coefficient. The introduction of this regularization term is similar to introducing a smooth prior to the source reconstruction, thus avoiding overfitting \citep{warren2003semilinear}. Throughout the modelling we use a cross-like regularization ensuring a smooth behaviour of the likelihood function, which guarantees a robust estimation of the errors of model parameters (see \citealt{nightingale2024scanning} and appendix A of \citealt{He2024}).   

    In this work, we use the image with the foreground galaxies subtracted as the adapt image $d_j$. When the weight map $w_j$ is calculated from the adapt image, two more parameters will be introduced: 
    \begin{equation}
    \label{eq:hilbert_filling}
      w_j =
      \begin{cases} 
         \hat{d}_j^{\omega_p}, & \text{if } w_j > \omega_f, \\ 
         \omega_f, & \text{if } w_j \leq \omega_f.
      \end{cases}
    \end{equation}
    where $\hat{d}_j = |d_j|/\max(d_j)$, $\omega_f$ is the weight floor, and $\omega_p$ is the weight power. The effect of the weight power is similar to increasing the contrast, while the weight floor ensures that the area around the arc has the lowest but not zero degree of grid point distribution.

    It is worth mentioning that the number of parameters for this brightness-adaptive source is four: $\omega_f$, $\omega_p$, $I$, and $\lambda$. In actual tests, considering the computational efficiency and the accuracy of source reconstruction, the number of pixels $I$ is fixed at 1000.
    In addition, adaptive regularization is not employed in the fitting (which controls smoothness variations across reconstruction regions). Instead, we use a constant regularization and fix the regularization coefficient to a relatively large value ($\lambda=20.0$) in \textbf{Model~5} to balance the physicality of the source reconstruction and the fineness of the model image. Although this coefficient could be treated as a free parameter, the fitting process consistently favors smaller regularization to achieve a lower $\chi^2$, which leads to an unphysical source reconstruction with discontinuous structures. This issue likely originates from the oversimplification of the mass model (as discussed in \ref{subsubsec:brightness-adaptive}), making such a large regularization necessary to ensure the physicality of the source reconstruction.
   
    \subsection{Sampling and search chain}
    \label{subsec:sampling_and_search_chain}
    We use the nested sampling algorithm \textsc{Dynesty} \citep{speagle2020dynesty}, specifically its static sampler with random walk sampling to perform non-linear parameter sampling fitting of each model. This method is less likely to fall into local optima than traditional MCMC algorithms.
    In \textsc{PyAutoLens}, when the fitting does not involve inversion, the likelihood function of the sampling is defined as:
    \begin{equation}
      -2 \ln \mathcal{L} = \chi^2 + \sum_{j=1}^{J} \ln(2\pi\sigma_j^2)
    \end{equation}
    where $\chi^2$ is defined as:
    \begin{equation}
      \chi^2 = \sum_{j=1}^{J} \left( \frac{d_j - m_j}{\sigma_j} \right)^2
    \end{equation}
    where $m_j$ is the pixel value of the model image.
    
    In the subsequent pixelized background source stage, the semilinear inversion framework described in Section \ref{subsec:source_model} is used. \textsc{PyAutoLens} uses the Bayesian framework first defined by \citet{suyu2006bayesiana} where the likelihood function aims to maximize the Bayesian evidence $\epsilon$:
    \begin{equation}
      -2 \ln \epsilon = \chi^2 + \lambda\mathbf{s}^T \mathbf{H} \mathbf{s} + \ln \left[ \det (\mathbf{F} + \lambda\mathbf{H}) \right] - \ln \left[ \det( \lambda\mathbf{H}) \right] + \sum_{j=1}^{J} \ln(2\pi\sigma_j^2)
    \end{equation}
    The notion of the above equation comes from Eq. (5) of \citet{dye2008models}.
    
    \begin{table*}
      \caption{Description of the search chain}
      \label{tab:search_chain}
      \centering
      \begin{tabular}{c c c}
      \hline
      \multirow{2}{*}{Model} & \multicolumn{2}{c}{Search Chain} \\ \cline{2-3}
                           & Source & Mass Components \\
      \hline
      \textbf{Model~1}   & 1 S\'ersic & main halo + BGG + L1 + Shear \\
      \textbf{Model~2}   & 1 S\'ersic & main halo + BGG + L1 + Shear + 14 members \\
      \textbf{Model~3}   & 2 S\'ersic & main halo + BGG + L1 + Shear + 14 members \\
      \textbf{Model~4}   & 3 S\'ersic & main halo + BGG + L1 + Shear + 14 members \\
      \textbf{Model~5-0} & Delaunay & fixed mass from model 4 \\
      \textbf{Model~5-1} & fixed Delaunay & main halo + BGG + L1 + Shear + 14 members \\
      \hline
      \end{tabular}
    \end{table*}

    \textsc{PyAutoLens} provides standardized pipelines for automated modelling of galaxy-scale lenses, known as the \texttt{Source, Light, and Mass (SLaM)} pipelines, used in works such as \citet{Cao2022, etherington2022automated}. These pipelines link together a sequence of non-linear searches that progressively increase model complexity, improving both efficiency and reliability while avoiding convergence to local optima in high-dimensional parameter spaces. For the system CSWA19, we build on the SLaM pipelines to make them appropriate for a group-scale lens and implement the custom search chaining strategy outlined in Table~\ref{tab:search_chain}. The light from the foreground galaxies is assumed to have been cleanly subtracted in advance, so their light profile parameters are fixed and not refined in subsequent steps.

    \textbf{Model~1} begins with a simple single S\'ersic background source, modelling only the main mass components: the primary halo, BGG, L1 (near the centre), and external shear. \textbf{Model~2} builds on this by using the results of \textbf{Model~1} as priors and adding the mass models of the remaining 14 member galaxies via a scaling relation. \textbf{Model~3} introduces a second S\'ersic component for the background source, again using the previous model as a prior. \textbf{Model~4} adds a third S\'ersic component. Once the three-S\'ersic fit is complete, we consider \textbf{Model~4} to provide sufficiently accurate priors for the final stage: an adaptive pixelized source reconstruction using a Delaunay mesh in \textbf{Model~5}. This is split into two steps: in \textbf{Model~5-0}, the mass model is fixed and only the pixelized source is fitted, to optimize the adaptive brightness parameters $\omega_f$ and $\omega_p$; in \textbf{Model~5-1}, these source parameters are fixed, and the mass model is refined based on the priors from \textbf{Model~4}.

    We apply a circular mask with a radius of 9\farcs5, using only pixels within the mask as lensing constraints. Different weighting strategies are used throughout the modelling stages. In \textbf{Model~1} and \textbf{Model~2}, we identify the region potentially corresponding to the S2 image and artificially inflate its noise by a factor of $10^8$ in the noise map. This effectively removes it from the fit, allowing us to focus on modelling the extended structure of S1. From \textbf{Model~3} onward, we increase the weight of the S2 image by a factor of 3 (due to its lower brightness relative to S1) to balance the constraints between S1 and S2. However, we retain high noise around the G2 galaxy and its surroundings due to the lack of spectroscopic confirmation of G2’s mass contribution, and therefore exclude this region from the fit. Notably, while G2 and its surroundings are excluded in the noise map, only G2 itself is masked in the construction of the adapt image. This enables us to test whether the nearby arcs can still be reconstructed despite being downweighted in the fitting process.

\section{Result}
\label{sec:result}
\begin{figure}
  \centering
    \begin{subfigure}[t]{0.495\textwidth}
        \centering
        \includegraphics[width=\linewidth]{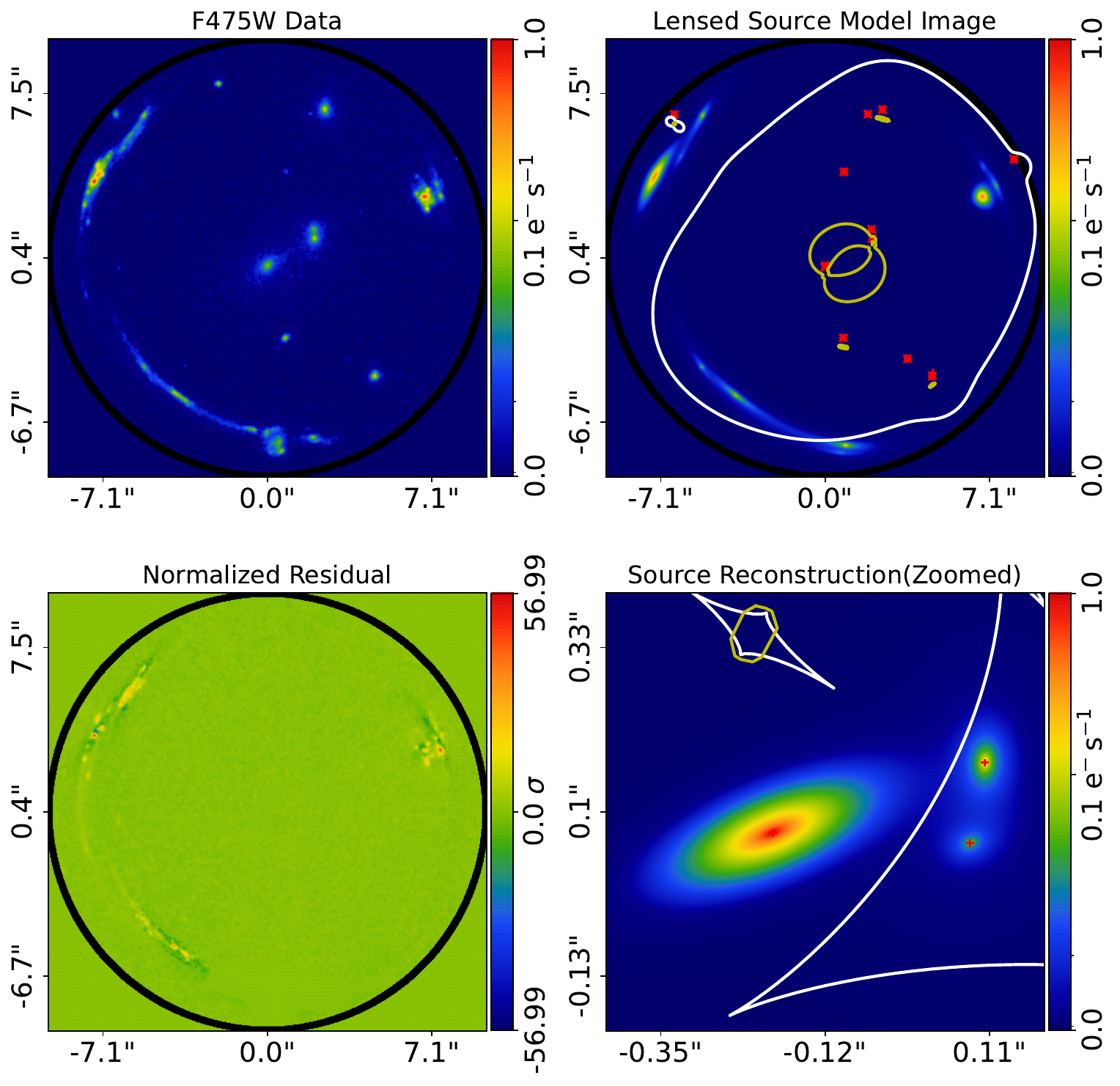}
        \caption{3-S\'ersic model ( \textbf{Model~4})}
        \label{subfig:3Sérsic}
    \end{subfigure}
    \hspace{-0.5em}
    \begin{subfigure}[t]{0.495\textwidth}
        \centering
        \includegraphics[width=\linewidth]{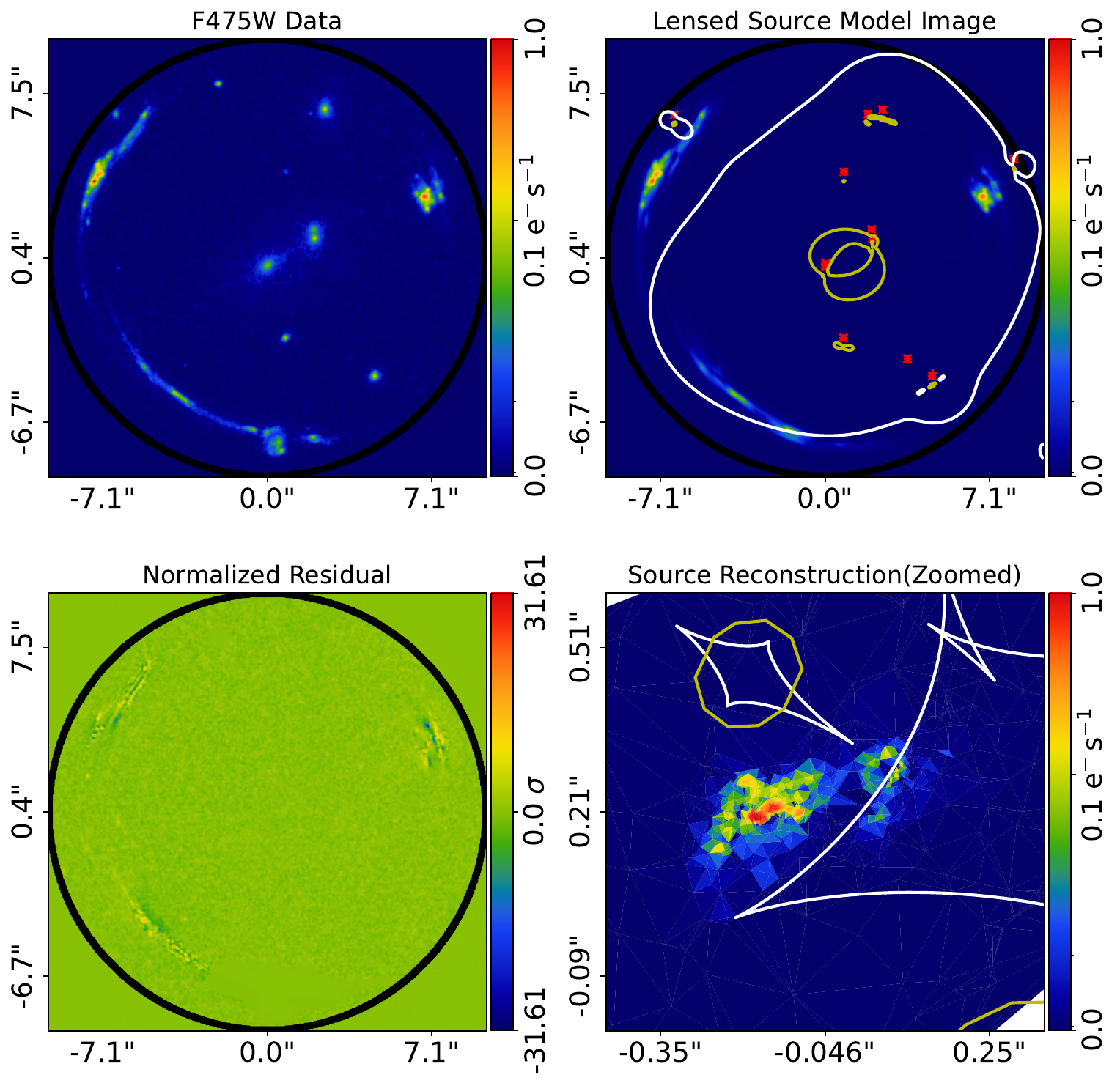}
        \caption{Adaptive-brightness pixelization model (\textbf{Model~5-1})}
        \label{subfig:pixelized}
    \end{subfigure}
  \caption{
  Comparison of lens modelling results.  The panels in the left part (\subref{subfig:3Sérsic}) show the 3-S\'ersic model results, and the panels in the right part (\subref{subfig:pixelized}) present the pixelized model results using a Delaunay mesh visualization for the source.
  Both parts share common features: 
  9\farcs5 radius circular mask (black circle) , tangential critical curves and caustics (white lines) and radial critical curves and caustics (yellow lines) 
  are overlaid, and red markers indicate mass component positions (including external L11$\sim$L15 member galaxies not visible in frame).
  }
  \label{fig:combined_results}
\end{figure}
    
In this section, we present the main modelling results of the above modelling process applied to CSWA19.
    
    \subsection{Modelling result from Parametric to Pixelized}
    \label{subsec:modelng_result}
    As shown in Table \ref{tab:search_chain}, we design a progressive search chain strategy to avoid introducing too many free parameters at once, and to gradually increase the complexity of the model. The parameter fitting results of each step are passed to the next step.
    We start with the simplest single S\'ersic background source model, increase model complexity, and finally switch to the pixelized source model. 
    In this process, the parameter space becomes more complex as the number of S\'ersic components describing the source increases.
    In the transition from \textbf{Model~1} to \textbf{Model~4}, the background source is described by 1 to 3 S\'ersic components, and in the transition from \textbf{Model~4} to \textbf{Model~5}, the background source is replaced by a pixelized model, so only the mass parameters are passed in this step. Finally, we take \textbf{Model~5} as the adopted model result.
    As shown in Fig. \ref{fig:combined_results}, we present the best-fitted modelling results of \textbf{Model~4} and \textbf{Model~5-1} in the search chain, which show the data image used as constraints, lensed source model image, normalized residual map and zoomed source reconstruction of their results. 
    
\begin{figure}
      \centering
      \includegraphics[width=0.6\textwidth]
      {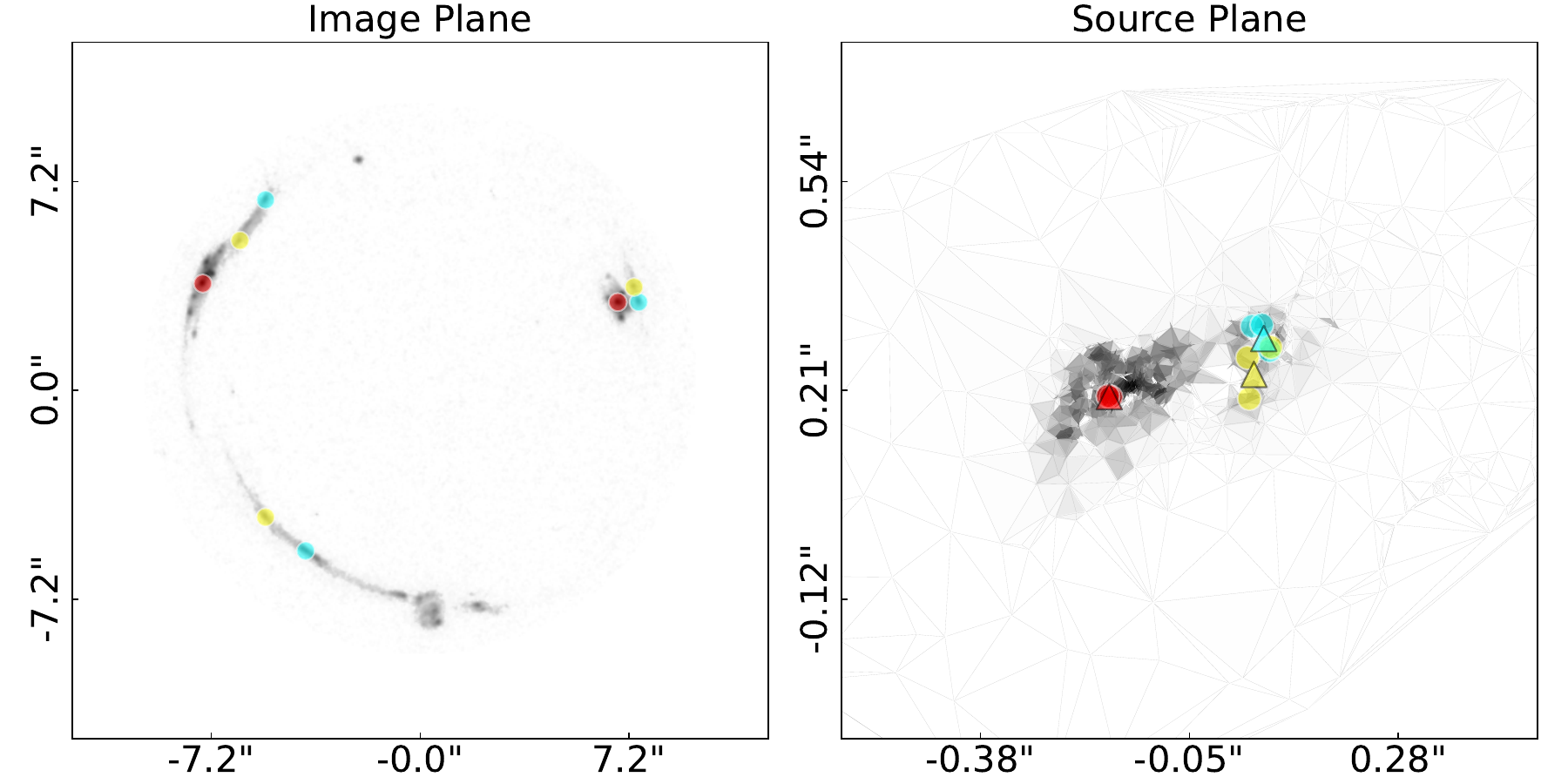} 
      \caption{Reconstructed positions on the source plane of multiple image point-like positions. The bottom layer of the left image is the image data with the lens light subtracted, with the brightest points of S1(a) and S1(b) marked in red circles, the brightest points of S2.1(a), S2.1(b), and S2.1(c) marked in cyan circles, and the brightest points of S2.2(a), S2.2(b), and S2.2(c) marked in yellow circles. Based on the best fitting results of \textbf{Model~5-1}, the right image shows the relative positions of these multiple images traced to the source plane, the triangles represent the weighted average centres of these coordinates, and the bottom layer is the source plane Delaunay triangle mesh source reconstruction.}
      \label{fig:traced_positions}
    \end{figure}

    We identify three groups of multiple images from visual inspection in Fig. \ref{fig:field_image}, although their positions are not used as lens constraints in our work. To verify the consistency between the extended source modelling result and the traced positions, we trace the peak points of these multiple images to the source plane based on the best-fitted results of \textbf{Model~5-1} (see Fig. \ref{fig:traced_positions}).
    The red circles mark the brightest points of S1(a) and S1(b), the cyan circles mark the brightest points of S2.1(a), S2.1(b), and S2.1(c), and the yellow circles mark the brightest points of S2.2(a), S2.2(b), and S2.2(c). We can see the relative positions of these points on the source plane, where the triangles represent the weighted average centres of these points (weighted by magnification). The average separation of the source plane positions of S1 from the weighted centre is $\sim 0\farcs003$, while for S2.1 it is $\sim 0\farcs02$ and for S2.2 it is $\sim 0\farcs04$. 
    
    Overall, our modelling result from both our 3-S\'ersic model and adaptive-brightness pixelization model successfully resolve the two background sources S1 and S2, which is consistent with the results of \citet{leethochawalit2016keck}. S1 forms a double image, corresponding to the S1(a) and S1(b) regions on the image plane, while S2 forms a quadruple image, corresponding to the S2(a), S2(b), and S2(c) regions on the image plane. The fourth image may be near G2, but we are not sure about the specific situation around it. 
    
\begin{figure}
      \centering
      \includegraphics[width=1.0\textwidth]
      {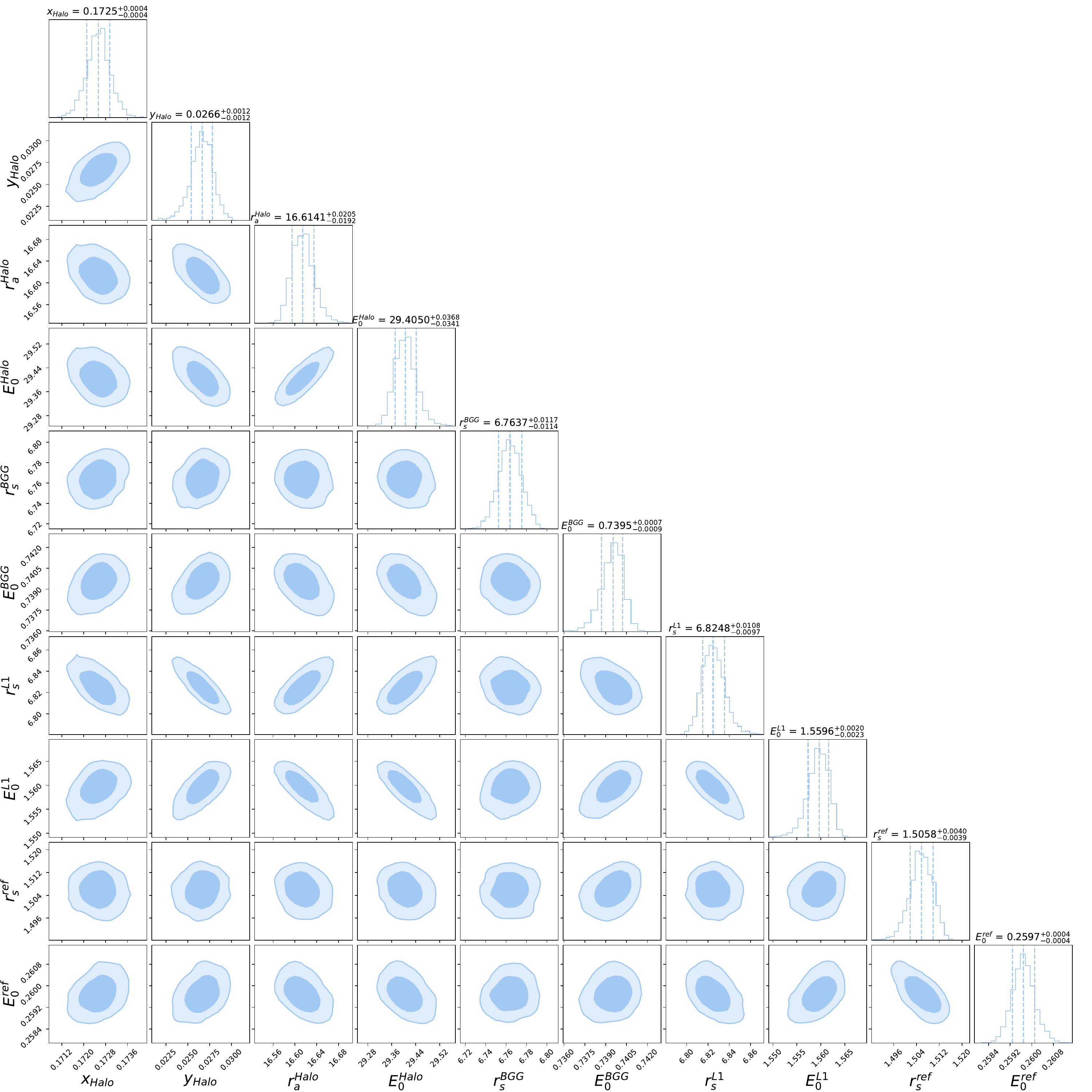} 
      \caption{The joint posterior probability distribution of the adaptive-brightness pixelization model fitting results of search chain \textbf{Model~5-1}, showing the mass component parameters of the BGG, the main halo, L1, and the reference galaxy that we think are important. The shadow area of the two-dimensional distribution shows the 68.3\% and 95.4\% confidence intervals, while the histograms on the diagonal show the marginalized distribution of each parameter, with the corresponding quantile range marked.}
      \label{fig:corner_plot}
    \end{figure}

    The best-fitted and marginalized results for some important parameters in \textbf{Model~5-1} with 1$\sigma$ error are provided in Table \ref{tab:fitting_result}, and their joint posterior probability distributions are shown in Fig. \ref{fig:corner_plot}.
    As a direct result of lens modelling, the brightness-adaptive image mesh grid construction is shown in Fig. \ref{fig:adaptive_grid}. Note that the construction of this grid is unique based on the adapt image and parameter values. The left image of Fig. \ref{fig:adaptive_grid} shows the data image with the lens light subtracted as the adapt image, with G2 masked, and the red circles show the adaptive grid points determined under the best-fitted results of \textbf{Model~5-0}. Under the mass profile of \textbf{Model~5-1}, these points can be traced to the source plane using the deflection angle to construct the Delaunay Triangulation.
    
    \begin{table}
      \centering
      \caption{The best-fit and marginalized values with 1$\sigma$ errors for the model parameters,  where the adaptive-brightness pixelization model with search chain \textbf{Model~5-1} is adopted.}
      \label{tab:fitting_result}
      \begin{tabular}{llcc}
        \toprule
        \multirow{2}{*}{Mass Component} & \multirow{2}{*}{Parameter} & \multicolumn{2}{c}{Fitting Result} \\
        \cmidrule(lr){3-4}
         & & Best-fit & Marginalized \\
        \midrule
        \multirow{2}{*}{BGG}  & $r_{s,\mathrm{BGG}}$ [${''}$] & 6.7617 & $6.7637^{+0.0117}_{-0.0114}$ \\
                              & $E_{0,\mathrm{BGG}}$ [${''}$] & 0.7395 & $0.7395^{+0.0007}_{-0.0009}$ \\
        \midrule
        \multirow{6}{*}{Halo} & $x_{\mathrm{Halo}}$ [${''}$]   & 0.1730 & $0.1725^{+0.0004}_{-0.0004}$ \\
                              & $y_{\mathrm{Halo}}$ [${''}$]   & 0.0269 & $0.0266^{+0.0012}_{-0.0012}$ \\
                              & $e_{1,\mathrm{Halo}}$          & 0.0831 & $0.0831^{+0.0002}_{-0.0002}$ \\
                              & $e_{2,\mathrm{Halo}}$          & -0.0120 & $-0.0116^{+0.0002}_{-0.0003}$ \\
                              & $r_{a,\mathrm{Halo}}$ [${''}$] & 16.6137 & $16.6141^{+0.0205}_{-0.0192}$ \\
                              & $E_{0,\mathrm{Halo}}$ [${''}$] & 29.4047 & $29.4050^{+0.0368}_{-0.0341}$ \\
        \midrule
        \multirow{2}{*}{L1}   & $r_{s,\mathrm{L1}}$ [${''}$] & 6.8246 & $6.8248^{+0.0108}_{-0.0097}$ \\
                              & $E_{0,\mathrm{L1}}$ [${''}$] & 1.5597 & $1.5596^{+0.0020}_{-0.0023}$ \\
        \midrule
        \multirow{2}{*}{Members} & $r_{s,\mathrm{ref}}$ [${''}$] & 1.5048 & $1.5058^{+0.0040}_{-0.0039}$ \\
                                 & $E_{0,\mathrm{ref}}$ [${''}$] & 0.2599 & $0.2597^{+0.0004}_{-0.0004}$ \\
        \midrule
        \multirow{2}{*}{External Shear} & $\gamma_{\mathrm{ext},1}$       & -0.0203 & $-0.0201^{+0.0001}_{-0.0001}$ \\
                                        & $\gamma_{\mathrm{ext},2}$       & 0.0119  & $0.0119^{+0.0001}_{-0.0001}$ \\
        \bottomrule
      \end{tabular}
    \end{table}

    \begin{figure}
      \centering
      \includegraphics[width=0.8\textwidth]
      {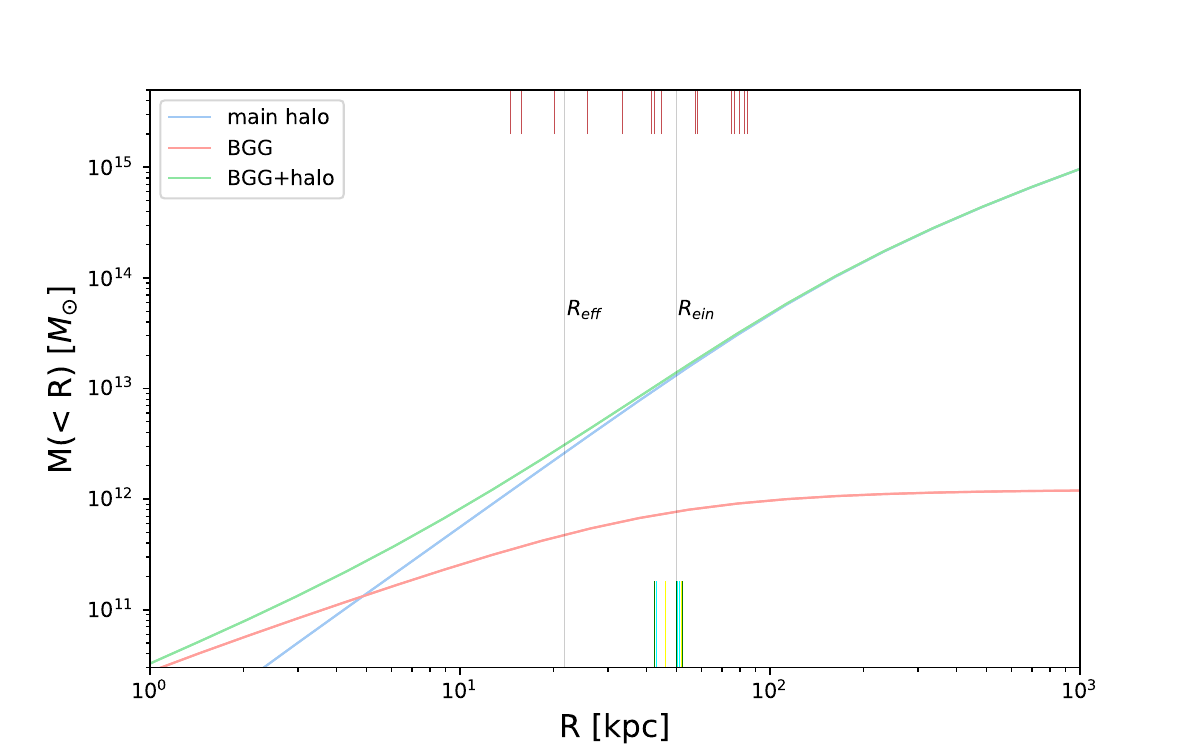} 
      \caption{The cumulative mass profiles within the radius $R$ (relative to the BGG centre) for BGG (red solid line), main halo (blue solid line), and total mass (green solid line). The 1-sigma error range is narrower than the line width due to small statistical errors. The vertical lines at the bottom represent the multiple image positions in Fig. \ref{fig:field_image}, with the same colour and label as in Fig. \ref{fig:field_image}. The red vertical lines at the top represent the positions of the member galaxies. We also use lines to mark the positions of BGG's effective radius $R_{\mathrm{eff}}=21.7$~kpc and the total Einstein radius $R_{\mathrm{ein}}=50.1$~kpc, $R_{\mathrm{eff}}$ is the best fitting value of BGG in Table \ref{tab:F160W_members_properties}, and $R_{\mathrm{ein}}$ is the Einstein radius of the best fitting results of \textbf{Model~5-1}.}
      \label{fig:enclosed_mass}
    \end{figure}
  
\section{Discussion}
\label{sec:discussion}

    \subsection{CSWA19 in the scientific context}
    \label{subsec:scientific_context}
    In addition to using CSWA19 as a testbed to verify the ability of \textsc{PyAutoLens} to model a group-scale system, we are also interested in the specific properties CSWA19 has as a particular example and what scientific insights it can provide.

        \subsubsection{Mass distribution of BGG and main halo}
        We calculate the cumulative mass distribution of BGG and the main halo as a function of the distance from the BGG centre in the \textbf{Model~5-1}, and the results are shown in Fig. \ref{fig:enclosed_mass}. The positions of the effective radius and the Einstein radius are marked with grey lines in the figure. 
        The effective radius of BGG, $R_{\mathrm{eff}}=21.7~\mathrm{kpc}$, is measured from the best fitted multi-S\'ersic model in Sec. \ref{subsec:foreground_galaxy_light} using the method outlined in \citet{simard2011catalog}. The Einstein radius is calculated from the equivalent circular radius of the area enclosed by the critical curve, which is defined as where the average surface density within the Einstein radius is equal to the critical surface density. The Einstein radius of the system is $\theta_E=8\farcs0653^{+0.0001}_{-0.0001}$, corresponding to a physical scale of $50.0840^{+0.0006}_{-0.0007}~\mathrm{kpc}$ on the image plane. 
        The vertical lines at the bottom of the image mark the positions of the multiple images S1, S2.1, and S2.2, and it can be seen that the positions of the images are consistent with the positions of the Einstein radius. The mass within the Einstein radius is $1.4065^{+0.0002}_{-0.0001} \times 10^{13}~M_\odot$ for total mass. 
        
        From the cumulative mass distribution (Fig. \ref{fig:enclosed_mass}), we find that the total mass at $R_{\mathrm{eff}}$ and $R_{\mathrm{ein}}$ is dominated by the main halo, whereas the BGG component dominants at smaller radii. However, caution is required, as the fitting may underestimate the BGG mass. This limitation arises because strong lensing alone cannot reliably distinguish between the dark matter and stellar components of BGGs. Without complementary methods like stellar dynamics to securely decompose these contributions, the uncertainty persists. Nonetheless, the total mass estimate remains reliable, and the Einstein radius and enclosed mass derived in this work are consistent with previous results from \citet{diehl2009sloana,stark2013cassowary}.
    
        \begin{figure}
          \centering
          \includegraphics[width=0.7\textwidth]
          {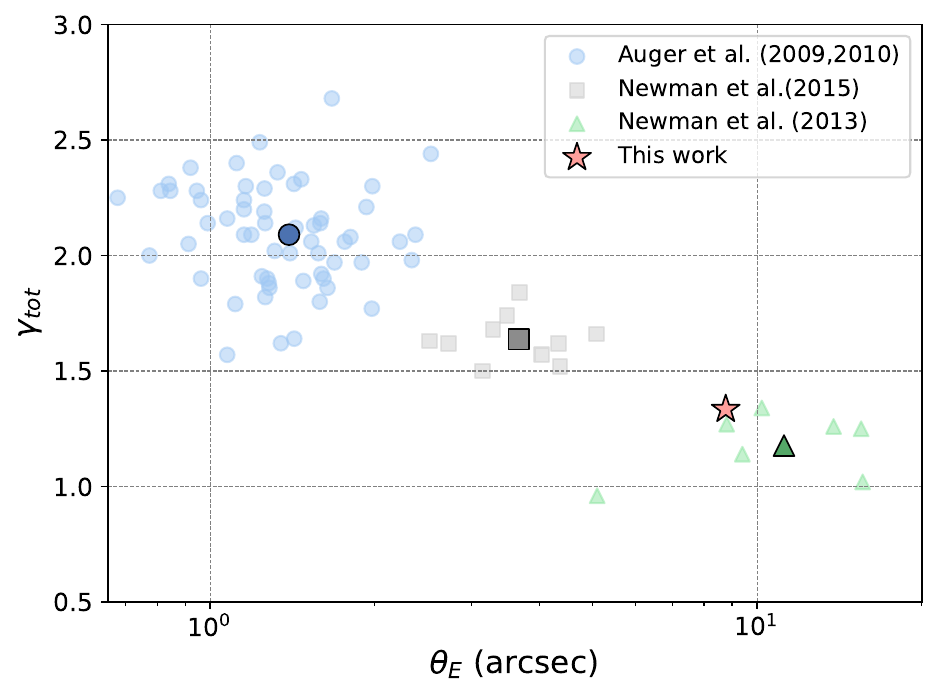} 
          \caption{Trends of the total mass density slope within $R_{\mathbf{eff}}$ as a function of $\theta_E$. These data points span a wide range of scales: galaxy-scale lenses from the SLACS survey \citep{auger2009sloan,auger2010sloanb} (blue circles), group-scale lenses in the work of \citet{newman2015luminous} (grey squares), and cluster-scale lenses in the work of \citet{newman2013densitya,newman2013densityb} (green triangles). The darker patterns corresponding to these different scale data points represent the average values of these data points. The results of CSWA19 are roughly located at the position of the red pentagram (with $\theta_{E}$ rescaled to a lensing ratio $D_{LS}/D_S=0.7$).}
          \label{fig:total_slope_relation}
        \end{figure}

        \subsubsection{Total density slope}
        The radial slope of the total mass density profile within the effective radius can be calculated by weighting the mass \citep{dutton2014bulgehalo}. Under the assumption of $\rho(r)\propto r^{-\gamma(r)}$ and the definition of $-\gamma(r)\equiv \mathbf{d}\log \rho / \mathbf{d}\log r$, the average slope $\tilde{\gamma}(r)$ within $r$ is defined as:
        \begin{equation}
          \tilde{\gamma}(r) = \frac{\int_0^r \gamma(x)\cdot 4\pi x^2\cdot \rho(x) \mathrm{d}x}{M(r)} = \frac{4\pi r^3 \cdot \tilde{\rho}(r)}{M(r)} - 3\cdot\frac{\rho(r)}{\tilde{\rho}(r)} = 3-3\frac{\rho(r)}{\tilde{\rho}(r)} = 3-\frac{4\pi r^3 \rho(r)}{M(r)}
        \end{equation}
        where $M(r)$ is the cumulative mass within the radius $r$: $M(r) = \frac{4}{3}\pi r^3 \cdot \tilde{\rho}(r)$, where $\tilde{\rho}(r)$ is the average mass density within $r$, and $\rho(r)$ is the density at $r$. The 3D density distribution of dPIE \citep{eliasdottir2007where} is:
        \begin{equation}
        \rho(r) =  \frac{\rho_0}{(1+r^2/r_a^2)(1+r^2/r_s^2)} = E_0 \frac{\Sigma_{\mathrm{crit}}}{2\pi}\frac{(r_s-r_a)(r_s+r_a)^2}{r_s\cdot(r_a^2+r^2)(r_s^2+r^2)}
        \end{equation}
        where $\Sigma_{\mathrm{crit}}$ is the critical surface density. Due to the lack of spectroscopic data and inclination angle measurements from stellar dynamics, we superimpose the density distributions of the BGG and main halo under this circular assumption. We then calculate the total density slope within $R_{\mathbf{eff}}$ to obtain $\tilde{\gamma}(R_{\mathbf{eff}}) = 1.3327^{+0.0005}_{-0.0006}$.
        
        Our result is contextualized with \citet{newman2015luminous}, as shown in Fig. \ref{fig:total_slope_relation}. They found that the total density slope becomes shallower in larger systems and emphasized the importance of group-scale lenses in mapping the full evolutionary trend of $\tilde{\gamma}$. This trend is crucial in understanding how baryonic matter affects the dark matter distribution at different scales. Our result is located at the position of the red pentagram, which is consistent with the trend. Although CSWA19 belongs to group scale for its number of member galaxies, it is closer to the results of the cluster-scale within the trend.

        \subsubsection{Source properties with big magnification}
        The total magnification $\mu_{\mathrm{tot}}$ of the source is defined as the flux ratio between the image plane and the source plane. We found $\mu_{\mathrm{tot,4}}=79.70^{+0.02}_{-0.03}$ in \textbf{Model~4} and $\mu_{\mathrm{tot,5-1}}=103.18^{+0.23}_{-0.19}$ in \textbf{Model~5-1}. The intrinsic magnitude of the background source in the F475W is thus $\sim$25~mag. 
        
        This magnification is significantly higher than previous estimates: $\mu = 6.5$ reported by \citet{stark2013cassowary}, based on an SIS model from \citet{diehl2009sloana} using ground-based SDSS imaging; and $\mu = 4.3$ from \citet{leethochawalit2016keck}, who used HST data but constrained their LTM model solely with multiple image positions.
        The increase in magnification can be attributed to the intrinsically compact size of the background source, with effective radii in the range of $0\farcs07\sim0\farcs28$ (as inferred from \textbf{Model~4}), which are significantly stretched and distorted to a much larger scale.
        Notably, source component S2 is positioned very close to the caustic curve (Fig. \ref{subfig:3Sérsic}) and even crosses it in the pixelized reconstruction (Fig. \ref{subfig:pixelized}). Previous models, constrained primarily by multiple image positions and based on the low-resolution SDSS imaging, likely underestimated the magnification. Our results highlight the power of extended source modelling in revealing highly magnified, small-scale structures that may be missed by traditional methods.

    \subsection{Brightness-adaptive pixelization as a paradigm shift}
    \label{subsec:pixelization_shift}
        
        \subsubsection{Challenges in Group-Scale Lens Modelling}
        The modelling of group-scale strong lenses presents unique challenges compared to galaxy-scale or cluster-scale systems. Unlike cluster-scale lenses, which often exhibit multiple independent lensed images from different background sources, group-scale lenses typically have fewer multiply imaged sources and lack complete spectroscopic confirmation.
        In Section \ref{sec:result}, we attempt to trace the brightest positions of multiple images back to the source plane despite the absence of spectroscopic confirmation (see Fig.~\ref{fig:traced_positions}). We can clearly distinguish the positions of S1 and S2 as two sources on the source plane, but distinguishing S2.1 and S2.2 becomes difficult due to the larger separations (especially S2.2). This is evidence that selecting multiple image positions in the image plane can be inaccurate for group scale systems with large magnifications and extended arcs. It also reinforces the limitation of using point-like positions as constraints for group-scale lensing models, highlighting the reliability of extended image constraints \citep{xie2024curling}.
        
        Using extended image pixels as constraints offers a significant advantage over traditional point-like constraints. As demonstrated in previous works \citep{wang2022constraining,urcelay2024compact,acebron2024next,xie2024curling}, pixel-based constraints yield narrower statistical errors (see our result in Table \ref{tab:fitting_result} and Fig. \ref{fig:corner_plot}) compared to position-based constraints (refer to Figure~8 in \citet{wang2022constraining}'s works, Figure~7 in \citet{acebron2024next}'s work and Figure~3 in \citet{xie2024curling}'s work). This improvement arises from the substantial increase in the number of independent constraints when using extended images.
        
        Previous studies have employed either parametric light profiles \citep{urcelay2024compact,xie2024curling} or pixelized reconstructions \citep{wang2022constraining,acebron2024next}. 
        While parametric models (e.g., the S\'ersic profile) struggle to represent complex and irregular brightness distributions, pixelized models demonstrate significantly lower residuals (see Fig. \ref{fig:combined_results}), as parametric approaches often fail to reproduce fine-grained surface brightness features.
        Compared to purely parametric models, such as the 3-S\'ersic model, pixelized models require very few non-linear free parameters to describe the background source. Even though the linear inversion involves a large number of linear algebra operations, which increases computational cost during single-step calculations, fewer free parameters still speeds up the overall convergence time. Ideally, for models with more nonlinear parameters, whose non-linear sampling becomes the dominant factor, the efficiency is higher with pixelized modelling.
    
        \subsubsection{Brightness-Adaptive Pixelization as a Solution}
        \label{subsubsec:brightness-adaptive}
        We now assess the advantages of adapting source reconstruction to the brightness distribution in group-scale lenses, rather than to the lensing magnification as done in many previous methods. Magnification-based approaches typically place dense, uniform grids across the image plane to ensure accurate reconstructions. However, in systems like CSWA19—where the lensed source spans a wide circular region with an $8\arcsec$ radius—a large number of source pixels are required to capture the full emission. If each image pixel were mapped to a unique source pixel, the total would exceed 100,000, making such reconstructions computationally prohibitive.
        
        The brightness-adaptive pixelization technique (Section \ref{subsec:source_model}) addresses this challenge by adjusting the grid density based on image brightness. This significantly reduces the number of source pixels and the dimensionality of the inversion matrices (Eqs. \ref{eq:Chi2_for_inversion}, \ref{eq:inversion_with_reg}). In our analysis, just $I = 1000$ source pixels are sufficient to model the brightness distribution, compared to $J = 177{,}144$ image pixels used for residual calculations—yielding substantial computational savings without compromising reconstruction fidelity.
        
        These gains are enabled by a Hilbert space-filling curve-based clustering algorithm (Fig. \ref{fig:adaptive_grid}), which concentrates source pixels in bright regions and sparsely samples faint ones. Although this method introduces two extra parameters ($\omega_f$, $\omega_p$ in Eq. \ref{eq:hilbert_filling}), our search chain (Section \ref{subsec:sampling_and_search_chain}) optimizes them separately from the lens model to avoid inflating parameter space. Our results show that brightness-adaptive pixelization is a scalable and efficient technique for group-scale lensing studies.
        
        \textbf{Model~5-1} delivers a physically meaningful source reconstruction, resolving a merging galaxy pair and spiral-arm-like features. However, the normalized residuals remain high (up to $\pm30\sigma$), indicating that finer structures are not fully recovered. We attribute this to an overly simplistic lens mass model. Improvements likely require both a more complex lens galaxy parameterization (e.g., using multipoles \citealt{amvrosiadis2024lopsidedness}) and relaxing the scaling relation assumptions for the member galaxies.
        
        Looking ahead, brightness-adaptive pixelization offers a promising route for future group-scale lens modelling. As datasets grow larger and higher in resolution, methods that efficiently reconstruct complex source structures while remaining computationally tractable will be essential. Our results underscore the need for pixelized techniques that adapt to brightness, paving the way for the next generation of lens modelling.

\section{Conclusion}
\label{sec:conclusion}
In this work, we fully utilised high-resolution HST/WFC3 imaging to model the galaxy group-scale strong lens system CASSOWARY 19 (CSWA19) at the pixel-level for the first time. We identified 16 member galaxies through red-sequence fitting (Fig. \ref{fig:red_sequence}) and subtracted their light contamination in the F475W and F160W bands (Fig. \ref{fig:combined_lens_light}). The foreground mass model—composed of a main halo, BGG, L1, and member galaxies in scaling relation—adopted pure dual pseudo-isothermal elliptical (dPIE) profiles. The lensed source was reconstructed using \textsc{PyAutoLens}'s brightness-adaptive pixelization technique, which distributes source pixels along Hilbert space-filling curves (Fig. \ref{fig:adaptive_grid}). This method enabled efficient reconstruction of the total 177,144 HST pixels (pixel scale=0\farcs04) in the WFC3/F475W image with only 1000 adaptive source pixels. 

By progressively refining the model from multi-S\'ersic parameterization to adaptive pixelization (Fig. \ref{fig:combined_results}), we resolved the merging pair of background galaxies and measured a total magnification of $\mu_{tot}=103.18^{+0.23}_{-0.19}$—significantly exceeding previous estimates—which further demonstrates the advantage of extended image modelling over traditional point-position-based methods. The total mass within the Einstein radius was determined to be $M_{\theta_E}=1.4065^{+0.0002}_{-0.0001} \times 10^{13}~M_\odot$, with an Einstein radius of $\theta_E=8\farcs0653^{+0.0001}_{-0.0001}$. Additionally, the total mass density slope within the effective radius was found to be $\tilde{\gamma}(R_{\mathbf{eff}})=1.3327^{+0.0005}_{-0.0006}$, consistent with the $\gamma_{tot}$-$\theta_E$ correlation trend presented in Fig. \ref{fig:total_slope_relation}. 

Future studies would benefit from integral field unit (IFU) spectroscopy to resolve ambiguities in member galaxy identification (e.g., G2) and enable stellar-dynamical modelling of the BGG. While our current residuals are likely dominated by systematic uncertainties in the simplified mass model, advancements in computational frameworks—such as GPU-accelerated optimization with JAX \citep{bradbury2018jax}—could enable more flexible parameterizations. Currently, extended source modelling of group- and cluster-scale lenses is emerging as a transformative approach \citep{acebron2024next,urcelay2024compact}. Upcoming next-generation surveys like LSST, Euclid, and CSST will discover large samples of group-scale lenses, for which our methodology provides a scalable template to exploit high-resolution imaging. By integrating efficient modelling strategies with adaptive pixelization, these datasets will enable statistical studies of density slopes across mass scales, thus enhancing our understanding of the interplay between baryonic and dark matter at various scales.
    
\section*{Acknowledgements}
This work has been supported by the National Key Research and Development Program of China (Nos. 2023YFB3002501 and 2022YFA1602903), the China Manned Space Project (No. CMS-CSST-2021-A01), and the National Natural Science Foundation of China (No. 11988101). L.W. acknowledges the support from the GHfund A (Nos. 202302017475 and 202407017555).
We thank Andrew B. Newman (Carnegie Observatories \& University of Southern California) for kindly providing the cluster-scale Einstein radii and normalization details critical to our comparative analysis.
In this study, a cluster is used with the SIMT accelerator made in China. The cluster includes many nodes each containing two CPUs and eight accelerators. The accelerator adopts a GPU-like architecture consisting of a 64 GB HBM2 device memory and many compute units. Accelerators connected to CPUs with PCI-E, the peak bandwidth of the data transcription between main memory and device memory is 64 GB/s.

\bibliographystyle{raa}
\bibliography{bibtex}

\begin{thebibliography}{133}
\providecommand\natexlab[1]{#1}
\providecommand\JournalTitle[1]{#1}

\bibitem[Acebron {et~al.}(2024)]{acebron2024next}
Acebron, A., Grillo, C., Suyu, S.~H., {et~al.} 2024, The next Step in Galaxy Cluster Strong Lensing: Modeling the Surface Brightness of Multiply-Imaged Sources, arXiv:2410.01883

\bibitem[Amvrosiadis {et~al.}(2024)]{amvrosiadis2024lopsidedness}
Amvrosiadis, A., Nightingale, J.~W., He, Q., {et~al.} 2024, Lopsidedness in {{Early-Type Galaxies}}: The Role of the \$m=1\$ Multipole in {{Isophote Fitting}} and {{Strong Lens Modelling}}

\bibitem[Auger {et~al.}(2009{\natexlab{a}})]{auger2009sloana}
Auger, M.~W., Treu, T., Bolton, A.~S., {et~al.} 2009{\natexlab{a}}, The Astrophysical Journal, 705, 1099

\bibitem[Auger {et~al.}(2009{\natexlab{b}})]{auger2009sloan}
Auger, M.~W., Treu, T., Bolton, A.~S., {et~al.} 2009{\natexlab{b}}, The Astrophysical Journal, 705, 1099

\bibitem[Auger {et~al.}(2010{\natexlab{a}})]{auger2010sloanb}
Auger, M.~W., Treu, T., Bolton, A.~S., {et~al.} 2010{\natexlab{a}}, The Astrophysical Journal, 724, 511

\bibitem[Auger {et~al.}(2010{\natexlab{b}})]{auger2010sloan}
Auger, M.~W., Treu, T., Bolton, A.~S., {et~al.} 2010{\natexlab{b}}, The Astrophysical Journal, 724, 511

\bibitem[{Avila} {et~al.}(2012)]{astrodrizzle}
{Avila}, R.~J., {Hack}, W.~J., \& {STScI AstroDrizzle Team}. 2012, in American Astronomical Society Meeting Abstracts, Vol. 220, American Astronomical Society Meeting Abstracts \#220, 135.13

\bibitem[Belokurov {et~al.}(2009)]{belokurov2009two}
Belokurov, V., Evans, N.~W., Hewett, P.~C., {et~al.} 2009, Monthly Notices of the Royal Astronomical Society, 392, 104

\bibitem[Bender {et~al.}(1992)]{bender1992dynamically}
Bender, R., Burstein, D., \& Faber, S.~M. 1992, The Astrophysical Journal, 399, 462

\bibitem[Bergamini {et~al.}(2019)]{bergamini2019enhanced}
Bergamini, P., Rosati, P., Mercurio, A., {et~al.} 2019, Astronomy \& Astrophysics, 631, A130

\bibitem[Bergamini {et~al.}(2021)]{bergamini2021new}
Bergamini, P., Rosati, P., Vanzella, E., {et~al.} 2021, Astronomy \& Astrophysics, 645, A140

\bibitem[Bertin \& Arnouts(1996)]{sextractor}
Bertin, E., \& Arnouts, S. 1996, Astronomy and Astrophysics Supplement Series, 117, 393

\bibitem[{Biesiada}(2006)]{Biesiada2006PhRvD}
{Biesiada}, M. 2006, \prd, 73, 023006

\bibitem[{Birrer} {et~al.}(2022)]{psfr}
{Birrer}, S., {Bhamre}, V., {Nierenberg}, A., {Yang}, L., \& {Van de Vyvere}, L. 2022, {PSFr: Point Spread Function reconstruction}, Astrophysics Source Code Library, record ascl:2210.005

\bibitem[Bower {et~al.}(1992)]{bower1992precision}
Bower, R.~G., Lucey, J.~R., \& Ellis, R.~S. 1992, Monthly Notices of the Royal Astronomical Society, 254, 589

\bibitem[Bradbury {et~al.}(2018)]{bradbury2018jax}
Bradbury, J., Frostig, R., Hawkins, P., {et~al.} 2018, {JAX}: composable transformations of {P}ython+{N}um{P}y programs

\bibitem[Bradley {et~al.}(2023)]{photutils_1_10_0}
Bradley, L., Sipőcz, B., Robitaille, T., {et~al.} 2023, astropy/photutils: 1.10.0

\bibitem[Bro \& De~Jong(1997)]{bro1997fast}
Bro, R., \& De~Jong, S. 1997, Journal of Chemometrics: A Journal of the Chemometrics Society, 11, 393

\bibitem[Caminha {et~al.}(2019)]{caminha2019strong}
Caminha, G.~B., Rosati, P., Grillo, C., {et~al.} 2019, Astronomy \& Astrophysics, 632, A36

\bibitem[Cao {et~al.}(2012)]{cao2012constraints}
Cao, S., Pan, Y., Biesiada, M., Godlowski, W., \& Zhu, Z.-H. 2012, Journal of Cosmology and Astroparticle Physics, 2012, 016

\bibitem[{Cao} \& {Zhu}(2012)]{Cao2012AA}
{Cao}, S., \& {Zhu}, Z.-H. 2012, \aap, 538, A43

\bibitem[Cao {et~al.}(2025)]{cao2025csst}
Cao, X., Li, R., Li, N., {et~al.} 2025, {{CSST Strong Lensing Preparation}}: {{Fast Modeling}} of {{Galaxy-Galaxy Strong Lenses}} in the {{Big Data Era}}, arXiv:2503.08586

\bibitem[{Cao} {et~al.}(2022)]{Cao2022}
{Cao}, X., {Li}, R., {Nightingale}, J.~W., {et~al.} 2022, Research in Astronomy and Astrophysics, 22, 025014

\bibitem[Cava {et~al.}(2018)]{cava2018naturea}
Cava, A., Schaerer, D., Richard, J., {et~al.} 2018, Nature Astronomy, 2, 76

\bibitem[{Chen} {et~al.}(2019)]{Chen2019MNRAS}
{Chen}, Y., {Li}, R., {Shu}, Y., \& {Cao}, X. 2019, \mnras, 488, 3745

\bibitem[Chiriv{\`i} {et~al.}(2018)]{chirivi2018macs}
Chiriv{\`i}, G., Suyu, S.~H., Grillo, C., {et~al.} 2018, Astronomy \& Astrophysics, 614, A8

\bibitem[Ciotti \& Bertin(1999)]{ciotti1999analytical}
Ciotti, L., \& Bertin, G. 1999, Analytical Properties of the {{R}}{\textasciicircum}(1/m) Luminosity Law, arXiv:astro-ph/9911078

\bibitem[Conroy {et~al.}(2009)]{conroy2009propagation}
Conroy, C., Gunn, J.~E., \& White, M. 2009, The Astrophysical Journal, 699, 486

\bibitem[Dalal \& Kochanek(2002{\natexlab{a}})]{dalal2002directa}
Dalal, N., \& Kochanek, C.~S. 2002{\natexlab{a}}, The Astrophysical Journal, 572, 25

\bibitem[Dalal \& Kochanek(2002{\natexlab{b}})]{dalal2002direct}
Dalal, N., \& Kochanek, C.~S. 2002{\natexlab{b}}, The Astrophysical Journal, 572, 25

\bibitem[{de Vaucouleurs}(1961)]{devaucouleurs1961integrated}
{de Vaucouleurs}, G. 1961, The Astrophysical Journal Supplement Series, 5, 233

\bibitem[Diehl {et~al.}(2009{\natexlab{a}})]{diehl2009sloan}
Diehl, H.~T., Allam, S.~S., Annis, J., {et~al.} 2009{\natexlab{a}}, The Astrophysical Journal, 707, 686

\bibitem[Diehl {et~al.}(2009{\natexlab{b}})]{diehl2009sloana}
Diehl, H.~T., Allam, S.~S., Annis, J., {et~al.} 2009{\natexlab{b}}, The Astrophysical Journal, 707, 686

\bibitem[{Du} {et~al.}(2023)]{Du2023ApJ}
{Du}, W., {Fu}, L., {Shu}, Y., {et~al.} 2023, \apj, 953, 189

\bibitem[{Du} {et~al.}(2020)]{Du2020ApJ}
{Du}, W., {Zhao}, G.-B., {Fan}, Z., {et~al.} 2020, \apj, 892, 62

\bibitem[Dunham {et~al.}(2019)]{dunham2019lens}
Dunham, S.~J., Sharon, K., Florian, M.~K., {et~al.} 2019, The Astrophysical Journal, 875, 18

\bibitem[Dutton \& Treu(2014)]{dutton2014bulgehalo}
Dutton, A.~A., \& Treu, T. 2014, Monthly Notices of the Royal Astronomical Society, 438, 3594

\bibitem[Dye {et~al.}(2008)]{dye2008models}
Dye, S., Evans, N.~W., Belokurov, V., Warren, S.~J., \& Hewett, P. 2008, Monthly Notices of the Royal Astronomical Society, 388, 384

\bibitem[Einstein(1936)]{einstein1936lenslike}
Einstein, A. 1936, Science, 84, 506

\bibitem[El{\'i}asd{\'o}ttir {et~al.}(2007)]{eliasdottir2007where}
El{\'i}asd{\'o}ttir, {\'A}., Limousin, M., Richard, J., {et~al.} 2007, Where Is the Matter in the {{Merging Cluster Abell}} 2218?, arXiv:0710.5636

\bibitem[Etherington {et~al.}(2022)]{etherington2022automated}
Etherington, A., Nightingale, J.~W., Massey, R., {et~al.} 2022, Monthly Notices of the Royal Astronomical Society, 517, 3275

\bibitem[Faber {et~al.}(1987)]{faber1987globala}
Faber, S.~M., Dressler, A., Davies, R.~L., {et~al.} 1987, in Nearly {{Normal Galaxies}}, ed. S.~M. Faber (New York, NY: Springer), 175

\bibitem[Fassnacht {et~al.}(2002)]{fassnacht2002determination}
Fassnacht, C.~D., Xanthopoulos, E., Koopmans, L. V.~E., \& Rusin, D. 2002, The Astrophysical Journal, 581, 823

\bibitem[Fohlmeister {et~al.}(2007)]{fohlmeister2007time}
Fohlmeister, J., Kochanek, C.~S., Falco, E.~E., {et~al.} 2007, The Astrophysical Journal, 662, 62

\bibitem[Gilman {et~al.}(2020)]{gilman2020warm}
Gilman, D., Birrer, S., Nierenberg, A., {et~al.} 2020, Monthly Notices of the Royal Astronomical Society, 491, 6077

\bibitem[Gladders {et~al.}(1998)]{gladders1998slope}
Gladders, M.~D., {Lopez-Cruz}, O., Yee, H. K.~C., \& Kodama, T. 1998, The {{Slope}} of the {{Cluster Elliptical Red Sequence}}: {{A Probe}} of {{Cluster Evolution}}, arXiv:astro-ph/9802167

\bibitem[Gladders \& Yee(2000)]{gladders2000new}
Gladders, M.~D., \& Yee, H. K.~C. 2000, The Astronomical Journal, 120, 2148

\bibitem[Golse \& Kneib(2002)]{golse2002pseudo}
Golse, G., \& Kneib, J.-P. 2002, Astronomy and Astrophysics, v.390, p.821-827 (2002), 390, 821

\bibitem[Grillo {et~al.}(2015)]{grillo2015clashvlt}
Grillo, C., Suyu, S.~H., Rosati, P., {et~al.} 2015, The Astrophysical Journal, 800, 38

\bibitem[Grillo {et~al.}(2016)]{grillo2016story}
Grillo, C., Karman, W., Suyu, S.~H., {et~al.} 2016, The Astrophysical Journal, 822, 78

\bibitem[{He} {et~al.}(2024)]{He2024}
{He}, Q., {Nightingale}, J.~W., {Amvrosiadis}, A., {et~al.} 2024, \mnras, 532, 2441

\bibitem[He {et~al.}(2025)]{he2025notb}
He, Q., Robertson, A., Nightingale, J.~W., {et~al.} 2025, Not so Dark, Not so Dense: An Alternative Explanation for the Lensing Subhalo in {{SDSSJ0946}}+1006, arXiv:2506.07978

\bibitem[Hezaveh {et~al.}(2016)]{hezaveh2016detection}
Hezaveh, Y.~D., Dalal, N., Marrone, D.~P., {et~al.} 2016, The Astrophysical Journal, 823, 37

\bibitem[Hilbert(1891)]{hilbert1891stetige}
Hilbert, D. 1891, Mathematische Annalen, 38, 459

\bibitem[Johnson {et~al.}(2017{\natexlab{a}})]{johnson2017stara}
Johnson, T.~L., Sharon, K., Gladders, M.~D., {et~al.} 2017{\natexlab{a}}, The Astrophysical Journal, 843, 78

\bibitem[Johnson {et~al.}(2017{\natexlab{b}})]{johnson2017star}
Johnson, T.~L., Rigby, J.~R., Sharon, K., {et~al.} 2017{\natexlab{b}}, The Astrophysical Journal, 843, L21

\bibitem[Jones {et~al.}(2018)]{jones2018dust}
Jones, T., Stark, D.~P., \& Ellis, R.~S. 2018, The Astrophysical Journal, 863, 191

\bibitem[Jones {et~al.}(2015)]{jones2015grism}
Jones, T., Wang, X., Schmidt, K.~B., {et~al.} 2015, The Astronomical Journal, 149, 107

\bibitem[Jullo {et~al.}(2007)]{jullo2007bayesian}
Jullo, E., Kneib, J.-P., Limousin, M., {et~al.} 2007, New Journal of Physics, 9, 447

\bibitem[Kassiola \& Kovner(1993{\natexlab{a}})]{kassiola1993elliptica}
Kassiola, A., \& Kovner, I. 1993{\natexlab{a}}, The Astrophysical Journal, 417, 450

\bibitem[Kassiola \& Kovner(1993{\natexlab{b}})]{kassiola1993elliptic}
Kassiola, A., \& Kovner, I. 1993{\natexlab{b}}, The Astrophysical Journal, 417, 450

\bibitem[Kelly {et~al.}(2015)]{kelly2015multiple}
Kelly, P.~L., Rodney, S.~A., Treu, T., {et~al.} 2015, Science, 347, 1123

\bibitem[Kelly {et~al.}(2016)]{kelly2016deja}
Kelly, P.~L., Rodney, S.~A., Treu, T., {et~al.} 2016, The Astrophysical Journal, 819, L8

\bibitem[Kneib {et~al.}(1996)]{kneib1996hubble}
Kneib, J.-P., Ellis, R.~S., Smail, I., Couch, W.~J., \& Sharples, R.~M. 1996, The Astrophysical Journal, 471, 643

\bibitem[Kovner(1987)]{kovner1987marginal}
Kovner, I. 1987, Nature, Volume 325, Issue 6104, pp. 507-509 (1987)., 325, 507

\bibitem[Kubo {et~al.}(2009)]{kubo2009sloan}
Kubo, J.~M., Allam, S.~S., Annis, J., {et~al.} 2009, The Astrophysical Journal, 696, L61

\bibitem[{Lange} {et~al.}(2024)]{Lange2025}
{Lange}, S.~C., {Amvrosiadis}, A., {Nightingale}, J.~W., {et~al.} 2024, arXiv e-prints, arXiv:2410.12987

\bibitem[Leethochawalit {et~al.}(2016)]{leethochawalit2016keck}
Leethochawalit, N., Jones, T.~A., Ellis, R.~S., {et~al.} 2016, The Astrophysical Journal, 820, 84

\bibitem[{Li} \& {Chen}(2023)]{Li2023PDU}
{Li}, H., \& {Chen}, Y. 2023, Physics of the Dark Universe, 41, 101234

\bibitem[Limousin {et~al.}(2007{\natexlab{a}})]{limousin2007truncation}
Limousin, M., Kneib, J.~P., Bardeau, S., {et~al.} 2007{\natexlab{a}}, Astronomy and Astrophysics, 461, 881

\bibitem[Limousin {et~al.}(2009{\natexlab{a}})]{limousin2009probing}
Limousin, M., {Sommer-Larsen}, J., Natarajan, P., \& {Milvang-Jensen}, B. 2009{\natexlab{a}}, The Astrophysical Journal, 696, 1771

\bibitem[Limousin {et~al.}(2007{\natexlab{b}})]{limousin2007combining}
Limousin, M., Richard, J., Jullo, E., {et~al.} 2007{\natexlab{b}}, The Astrophysical Journal, 668, 643

\bibitem[Limousin {et~al.}(2009{\natexlab{b}})]{limousin2009newa}
Limousin, M., Cabanac, R., Gavazzi, R., {et~al.} 2009{\natexlab{b}}, Astronomy \& Astrophysics, 502, 445

\bibitem[Mainali {et~al.}(2023)]{mainali2023spectroscopy}
Mainali, R., Stark, D.~P., Jones, T., {et~al.} 2023, Monthly Notices of the Royal Astronomical Society, 520, 4037

\bibitem[Martis {et~al.}(2024)]{martis2024modelling}
Martis, N.~S., Sarrouh, G. T.~E., Willott, C.~J., {et~al.} 2024, Modelling and {{Subtracting Diffuse Cluster Light}} in {{JWST Images}}: {{A Relation}} between the {{Spatial Distribution}} of {{Globular Clusters}}, {{Dwarf Galaxies}}, and {{Intracluster Light}} in the {{Lensing Cluster SMACS}} 0723, arXiv:2401.01945

\bibitem[Meneghetti {et~al.}(2020)]{meneghetti2020excessa}
Meneghetti, M., Davoli, G., Bergamini, P., {et~al.} 2020, Science, 369, 1347

\bibitem[Me{\v s}tri{\'c} {et~al.}(2022)]{mestric2022exploring}
Me{\v s}tri{\'c}, U., Vanzella, E., Zanella, A., {et~al.} 2022, Monthly Notices of the Royal Astronomical Society, 516, 3532

\bibitem[Millon {et~al.}(2020)]{millon2020cosmograil}
Millon, M., Courbin, F., Bonvin, V., {et~al.} 2020, Astronomy and Astrophysics, 640, A105

\bibitem[More {et~al.}(2012)]{more2012cfhtlsstrong}
More, A., Cabanac, R., More, S., {et~al.} 2012, The Astrophysical Journal, 749, 38

\bibitem[Natarajan {et~al.}(2009)]{natarajan2009survival}
Natarajan, P., Kneib, J.-P., Smail, I., {et~al.} 2009, The Astrophysical Journal, 693, 970

\bibitem[Newman {et~al.}(2015)]{newman2015luminous}
Newman, A.~B., {Richard S. Ellis}, \& Treu, T. 2015, The Astrophysical Journal, 814, 26

\bibitem[Newman {et~al.}(2013{\natexlab{a}})]{newman2013densitya}
Newman, A.~B., Treu, T., Ellis, R.~S., \& Sand, D.~J. 2013{\natexlab{a}}, The Astrophysical Journal, 765, 25

\bibitem[Newman {et~al.}(2013{\natexlab{b}})]{newman2013densityb}
Newman, A.~B., Treu, T., Ellis, R.~S., {et~al.} 2013{\natexlab{b}}, The Astrophysical Journal, 765, 24

\bibitem[Newman {et~al.}(2013{\natexlab{c}})]{newman2013density}
Newman, A.~B., Treu, T., Ellis, R.~S., {et~al.} 2013{\natexlab{c}}, The Astrophysical Journal, 765, 24

\bibitem[{Nightingale} {et~al.}(2021{\natexlab{a}})]{pyautofit}
{Nightingale}, J., {Hayes}, R., \& {Griffiths}, M. 2021{\natexlab{a}}, The Journal of Open Source Software, 6, 2550

\bibitem[Nightingale \& Dye(2015)]{nightingale2015adaptive}
Nightingale, J.~W., \& Dye, S. 2015, Monthly Notices of the Royal Astronomical Society, 452, 2940

\bibitem[{Nightingale} {et~al.}(2018)]{autolens}
{Nightingale}, J.~W., {Dye}, S., \& {Massey}, R.~J. 2018, \mnras, 478, 4738

\bibitem[Nightingale {et~al.}(2024)]{nightingale2024scanning}
Nightingale, J.~W., He, Q., Cao, X., {et~al.} 2024, Monthly Notices of the Royal Astronomical Society, 527, 10480

\bibitem[{Nightingale} {et~al.}(2021{\natexlab{b}})]{pyautolens}
{Nightingale}, J., {Hayes}, R., {Kelly}, A., {et~al.} 2021{\natexlab{b}}, The Journal of Open Source Software, 6, 2825

\bibitem[Nightingale {et~al.}(2023)]{nightingale2023pyautogalaxy}
Nightingale, {\relax James}., Amvrosiadis, A., Hayes, R., {et~al.} 2023, The Journal of Open Source Software, 8, 4475

\bibitem[O'Donnell {et~al.}(2025)]{odonnell2025constraint}
O'Donnell, J.~H., Jeltema, T.~E., Roberts, M.~G., {et~al.} 2025, A {{Constraint}} on {{Dark Matter Self-Interaction}} from {{Combined Strong Lensing}} and {{Stellar Kinematics}} in {{MACS J0138-2155}}

\bibitem[Pascale {et~al.}(2024)]{pascale2024sn}
Pascale, M., Frye, B.~L., Pierel, J. D.~R., {et~al.} 2024, {{SN H0pe}}: {{The First Measurement}} of \${{H}}\_0\$ from a {{Multiply-Imaged Type Ia Supernova}}, {{Discovered}} by {{JWST}}, arXiv:2403.18902

\bibitem[{Planck Collaboration} {et~al.}(2016)]{planckcollaboration2016planck}
{Planck Collaboration}, Ade, P. A.~R., Aghanim, N., {et~al.} 2016, Astronomy and Astrophysics, 594, A13

\bibitem[Refsdal(1964)]{refsdal1964possibility}
Refsdal, S. 1964, Monthly Notices of the Royal Astronomical Society, 128, 307

\bibitem[Richard {et~al.}(2021)]{richard2021atlas}
Richard, J., Claeyssens, A., Lagattuta, D.~J., {et~al.} 2021, Astronomy \& Astrophysics, 646, A83

\bibitem[Rigby {et~al.}(2018)]{rigby2018magellan}
Rigby, J.~R., Bayliss, M.~B., Sharon, K., {et~al.} 2018, The Astronomical Journal, 155, 104

\bibitem[{Robitaille} {et~al.}(2020)]{reproject}
{Robitaille}, T., {Deil}, C., \& {Ginsburg}, A. 2020, {reproject: Python-based astronomical image reprojection}, Astrophysics Source Code Library, record ascl:2011.023

\bibitem[Schechter {et~al.}(1997)]{schechter1997quadruple}
Schechter, P.~L., Bailyn, C.~D., Barr, R., {et~al.} 1997, The Astrophysical Journal, 475, L85

\bibitem[Schuldt {et~al.}(2019{\natexlab{a}})]{schuldt2019innera}
Schuldt, S., Chiriv{\`i}, G., Suyu, S.~H., {et~al.} 2019{\natexlab{a}}, Astronomy and Astrophysics, 631, A40

\bibitem[Schuldt {et~al.}(2019{\natexlab{b}})]{schuldt2019inner}
Schuldt, S., Chiriv{\`i}, G., Suyu, S.~H., {et~al.} 2019{\natexlab{b}}, Astronomy \& Astrophysics, 631, A40

\bibitem[S{\'e}rsic(1963)]{sersic1963influence}
S{\'e}rsic, J.~L. 1963, Boletin de la Asociacion Argentina de Astronomia La Plata Argentina, 6, 41

\bibitem[Shajib(2019)]{shajib2019unified}
Shajib, A.~J. 2019, Monthly Notices of the Royal Astronomical Society, 488, 1387

\bibitem[Shajib {et~al.}(2023)]{shajib2023tdcosmo}
Shajib, A.~J., Mozumdar, P., Chen, G. C.~F., {et~al.} 2023, Astronomy and Astrophysics, 673, A9

\bibitem[Sharon {et~al.}(2022)]{sharon2022cosmic}
Sharon, K., Mahler, G., {Rivera-Thorsen}, T.~E., {et~al.} 2022, The Astrophysical Journal, 941, 203

\bibitem[Shu {et~al.}(2016)]{shu2016boss}
Shu, Y., Bolton, A.~S., Mao, S., {et~al.} 2016, The Astrophysical Journal, 833, 264

\bibitem[Shu {et~al.}(2017)]{shu2017sloan}
Shu, Y., Brownstein, J.~R., Bolton, A.~S., {et~al.} 2017, The Astrophysical Journal, 851, 48

\bibitem[Simard {et~al.}(2011)]{simard2011catalog}
Simard, L., Mendel, J.~T., Patton, D.~R., Ellison, S.~L., \& McConnachie, A.~W. 2011, The Astrophysical Journal Supplement Series, 196, 11

\bibitem[Speagle(2020)]{speagle2020dynesty}
Speagle, J.~S. 2020, Monthly Notices of the Royal Astronomical Society, 493, 3132

\bibitem[Stacey {et~al.}(2025)]{stacey2025nuclear}
Stacey, H.~R., Kaasinen, M., O'Riordan, C.~M., {et~al.} 2025, Astronomy and Astrophysics, 693, L17

\bibitem[Stark {et~al.}(2013)]{stark2013cassowary}
Stark, D.~P., Auger, M., Belokurov, V., {et~al.} 2013, Monthly Notices of the Royal Astronomical Society, 436, 1040

\bibitem[Stott {et~al.}(2009)]{stott2009evolution}
Stott, J.~P., Pimbblet, K.~A., Edge, A.~C., Smith, G.~P., \& Wardlow, J.~L. 2009, The Evolution of the Red Sequence Slope in Massive Galaxy Clusters, arXiv:0901.1227

\bibitem[Suyu \& Halkola(2010)]{suyu2010halos}
Suyu, S.~H., \& Halkola, A. 2010, Astronomy \& Astrophysics, 524, A94

\bibitem[Suyu {et~al.}(2010)]{suyu2010dissecting}
Suyu, S.~H., Marshall, P.~J., Auger, M.~W., {et~al.} 2010, The Astrophysical Journal, 711, 201

\bibitem[Suyu {et~al.}(2006{\natexlab{a}})]{suyu2006bayesian}
Suyu, S.~H., Marshall, P.~J., Hobson, M.~P., \& Blandford, R.~D. 2006{\natexlab{a}}, Monthly Notices of the Royal Astronomical Society, 371, 983

\bibitem[Suyu {et~al.}(2006{\natexlab{b}})]{suyu2006bayesiana}
Suyu, S.~H., Marshall, P.~J., Hobson, M.~P., \& Blandford, R.~D. 2006{\natexlab{b}}, Monthly Notices of the Royal Astronomical Society, 371, 983

\bibitem[{Turner} {et~al.}(1984)]{Turner1984ApJ}
{Turner}, E.~L., {Ostriker}, J.~P., \& {Gott}, III, J.~R. 1984, \apj, 284, 1

\bibitem[Urcelay {et~al.}(2024)]{urcelay2024compact}
Urcelay, F., Jullo, E., Barrientos, L.~F., Huang, X., \& Hernandez, J. 2024, A Compact Group Lens Modeled with {{GIGA-Lens}}: {{Enhanced}} Inference for Complex Systems, arXiv:2412.04567

\bibitem[Vegetti {et~al.}(2010{\natexlab{a}})]{vegetti2010quantifying}
Vegetti, S., Czoske, O., \& Koopmans, L. V.~E. 2010{\natexlab{a}}, Quantifying Dwarf Satellites through Gravitational Imaging: The Case of {{SDSS J120602}}.09+514229.5, arXiv:1002.4708

\bibitem[Vegetti \& Koopmans(2009)]{vegetti2009bayesian}
Vegetti, S., \& Koopmans, L. V.~E. 2009, Monthly Notices of the Royal Astronomical Society, 392, 945

\bibitem[Vegetti {et~al.}(2010{\natexlab{b}})]{vegetti2010detection}
Vegetti, S., Koopmans, L. V.~E., Bolton, A., Treu, T., \& Gavazzi, R. 2010{\natexlab{b}}, Monthly Notices of the Royal Astronomical Society, 408, 1969

\bibitem[{Wang} {et~al.}(2020)]{Wang2020ApJ}
{Wang}, B., {Qi}, J.-Z., {Zhang}, J.-F., \& {Zhang}, X. 2020, \apj, 898, 100

\bibitem[Wang {et~al.}(2024{\natexlab{a}})]{wang2024manga}
Wang, C., Li, R., Zhu, K., {et~al.} 2024{\natexlab{a}}, Monthly Notices of the Royal Astronomical Society, 527, 1580

\bibitem[Wang {et~al.}(2024{\natexlab{b}})]{wang2024stronglensing}
Wang, H., Canameras, R., Suyu, S.~H., {et~al.} 2024{\natexlab{b}}, Strong-Lensing and Kinematic Analysis of {{CASSOWARY}} 31: Can Strong Lensing Constrain the Masses of Multi-Plane Lenses?, arXiv:2404.13205

\bibitem[Wang {et~al.}(2022)]{wang2022constraining}
Wang, H., Ca{\~n}ameras, R., Caminha, G.~B., {et~al.} 2022, Astronomy \& Astrophysics, 668, A162

\bibitem[Wang {et~al.}(2025)]{wang2025measuring}
Wang, K., Cao, X., Li, R., {et~al.} 2025, Measuring the {{Stellar-to-Halo Mass Relation}} at \${\textbackslash}sim10{\textasciicircum}\{10\}\$ {{Solar}} Masses, Using Space-Based Imaging of Galaxy-Galaxy Strong Lenses, arXiv:2501.16139

\bibitem[Warren \& Dye(2003)]{warren2003semilinear}
Warren, S.~J., \& Dye, S. 2003, The Astrophysical Journal, 590, 673

\bibitem[{Wei} {et~al.}(2022)]{Wei2022ApJ}
{Wei}, J.-J., {Chen}, Y., {Cao}, S., \& {Wu}, X.-F. 2022, \apjl, 927, L1

\bibitem[Wetzel \& White(2010)]{wetzel2010what}
Wetzel, A.~R., \& White, M. 2010, Monthly Notices of the Royal Astronomical Society, 403, 1072

\bibitem[Wong {et~al.}(2020)]{wong2020h0licow}
Wong, K.~C., Suyu, S.~H., Chen, G. C.~F., {et~al.} 2020, Monthly Notices of the Royal Astronomical Society, 498, 1420

\bibitem[Wuyts {et~al.}(2014)]{wuyts2014magnified}
Wuyts, E., Rigby, J.~R., Gladders, M.~D., \& Sharon, K. 2014, The Astrophysical Journal, 781, 61

\bibitem[Xie {et~al.}(2024)]{xie2024curling}
Xie, Y., Shan, H., Li, N., {et~al.} 2024, {{CURLING}} - {{I}}. {{The Influence}} of {{Point-like Image Approximation}} on the {{Outcomes}} of {{Cluster Strong Lens Modeling}}, arXiv:2405.03135

\bibitem[Yuan {et~al.}(2011)]{yuan2011metallicity}
Yuan, T.~T., Kewley, L.~J., Swinbank, A.~M., Richard, J., \& Livermore, R.~C. 2011, The Astrophysical Journal, 732, L14

\bibitem[Zitrin {et~al.}(2015)]{zitrin2015hubble}
Zitrin, A., Fabris, A., Merten, J., {et~al.} 2015, The Astrophysical Journal, 801, 44

\end{thebibliography}

\end{document}